\documentclass[twocolumn, 10pt, letterpaper]{article}
\usepackage[pass, margin=1in]{geometry} %one inch margins
\usepackage{ibpsa} % defines look
\usepackage{pslatex}
\usepackage{achicago}
\usepackage{amsmath, amssymb}
\usepackage{graphicx}
\usepackage{fancyhdr}
\usepackage{ifthen}
\usepackage[dvipsnames]{xcolor}
\usepackage[textfont=it, labelfont=it]{caption}
\usepackage{makecell}
\usepackage{multirow}
\usepackage{gensymb}
\usepackage{xcolor}
\usepackage{subcaption}
\usepackage{algorithm}
\usepackage{textcomp}
\usepackage{float}
\usepackage{algpseudocode}
\usepackage{soul}
\usepackage{amsmath}
\usepackage[pdftex,hypertexnames=false,linktocpage=true]{hyperref}\hypersetup{colorlinks=false,linkcolor=blue,anchorcolor=blue,citecolor=blue,filecolor=blue,urlcolor=blue,bookmarksnumbered=true,pdfview=FitB}
\PassOptionsToPackage{hyphens}{url}
\renewcommand{\eqref}[1]{(\ref{#1})}

\begin{document}
\date{}

\fancypagestyle{empty}{%
  \fancyhf{}% Clear header/footer
  
  \vspace{0.5in}%
  \fancyhead[L]{
  \begin{tabular}{cc}
   \includegraphics [width=0.48in] {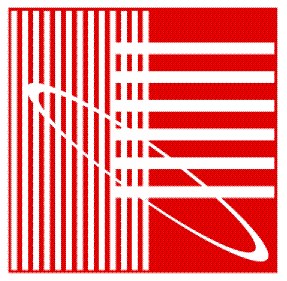} & \makecell[l]{\hspace{-0.6em}JOURNAL OF BUILDING PERFORMANCE SIMULATION \\  \vspace{.6em} \\ ~} \\
  \end{tabular}
  } %  logo and title
}

\ifthenelse{\value{page}=1}{\setlength{\textheight}{8.5in}}{\setlength{\textheight}{9in}}

\title{\vspace{-0.4in} \titlefont % do not change this line
    Optimizing Chilled Water Systems with Cooling Towers via Virtual Power Metrics and Extremum-Seeking Control\vspace{-0.2in} % do not change this line
}% do not change this line
% Dynamic Energy Optimization of Chilled Water Systems
% Real-Time Optimization for Energy Efficiency in Chilled Water Plants
% Optimizing Energy Use in Chilled Water Systems with Real-Time Control

\author{%
    \authorfont{~}\\% do not change this line
         % \authorfont{~}\\% 
         %     \authorfont{~}\\% 
         %         \authorfont{~}\\% 
    \authorfont{Min Gyung Yu$^1$,  Alex Vlachokostas$^1$, Karthik Devaprasad$^1$,    Matt Cornachione$^1$,}\\
    \authorfont{Stephanie Johnson$^1$, Tim A. Yoder$^1$, and Timothy I. Salsbury$^1$}\\
    \authorfont{$^1$Pacific Northwest National Laboratory, WA, USA}\\
    %The header consists of 10 lines with exactly 14 point spacing.
    \authorfont{~}\\ % used to add blank lines
    \authorfont{~}\\ % used to add blank lines
    \authorfont{~}\\ % used to add blank lines
    \authorfont{~}\\ % used to add blank lines
    %\authorfont{~}\\ % used to add blank lines
    %\authorfont{~}\\ % used to add blank lines
    % \authorfont{~}\\ % used to add blank lines
    \vspace{-0.55in} % do not change this line
}

\maketitle
\thispagestyle{empty}

% --------------------------------------------------------
\section{Abstract}
This paper presents an extremum seeking control (ESC) method for cooling tower fans to minimize overall power consumption of a chilled water plant system. Simulation studies across different climate locations demonstrate energy savings of approximately 15\% compared to conventional control during summer conditions. This paper also proposes a virtual power meter (VPM) to enable use of the strategy in systems that lack physical power meters.  Validation tests for the VPMs against physical meters showed good accuracy with a correlation of 96.11\% and a normalized error of 5.11\%. Coupled with the VPM, the proposed ESC control solution can be implemented on systems using typically available sensor measurements without the need for additional instrumentation. 
% key words - real time optimization, chiller water plant, vir. tual power meter, energy simulation, energy modeling
% --------------------------------------------------------
\section{Introduction}
\label{sect:intro}
% \subsection{Background and Motivation}
% \textcolor{red}{introduction to chilled water plants and energy challenges (1) importance of chilled water plants in HVAC systems (2) overview of energy consumption patterns in chilled water systems (3) relevance of energy efficiency and real-time optimization}
The energy consumption in buildings, particularly in the form of electricity, has become a significant contributor to global energy use. In fact, buildings account for 30\% of global final energy consumption, with electricity alone comprising about 35\% of the energy used within buildings in 2022 \cite{iea2022}. The electricity demand has been rising annually, and a major driver of this increase is space cooling. Since 1990, energy consumption for space cooling has more than tripled, with an average annual growth rate of 4\%, which is twice as fast as the increase in energy demand for water heating \cite{iea2022}.

Commercial buildings, in particular, use a substantial amount of electricity, consuming 775 TBtu in the U.S. --three times more than the 250 TBtu used for natural gas \cite{CBECS}. Buildings located in warmer climates show significantly higher electricity consumption for cooling compared to buildings in cooler regions.

Chillers are responsible for a large portion of the electricity used for space cooling. Although only 3\% of buildings use chillers, they account for 19\% of total floor space cooling \cite{CBECS}. Thus, the operation of these chillers can be particularly energy-intensive, contributing up to 40\% of a building's total electricity use \cite{Wei}. 

Given the high energy demands, especially for cooling, it is crucial to develop strategies to reduce energy use in electricity-driven chiller systems used in commercial buildings. This paper focuses on control-based strategies that optimize energy use without requiring additional equipment or instrumentation.

Chiller plant system typically consists of three main components: chillers, cooling towers, and pumps. These components work together to provide efficient cooling. Chillers are responsible for cooling the water, cooling towers reject heat to the outside, and pumps circulate the water through the system.

Optimizing energy consumption in a chiller plant system involves balancing the energy use of these components. As cooling towers remove heat from the condenser loop, the workload on the chiller is reduced. However, this process creates a trade-off between the energy used by the condenser water loop and the chiller system itself. In other words, while increasing the cooling tower fan speed can improve heat rejection and reduce the chiller's workload, it also increases fan energy consumption. Finding the right balance between these two competing energy uses is key in this optimization problem. 

Conventionally, these chiller plant systems rely on feedback-based control strategies that regulate chilled water temperature setpoints and maintain condenser water temperature. While these methods ensure stable operation, they do not directly address energy use. %Additionally, these systems often depend on sensor measurements, which can be unstable.

% \subsection{Research Objectives and Contributions}
% \textcolorsensor measurements, which can be unstable.{red}{Mention that the real-world implementation of the control system is the ultimate goal, but that this paper focuses on simulation-based validation and demonstration of the virtual power meter?}

In this paper, we propose a real-time controller to minimize the overall energy use of the chiller plant by dynamically adjusting the cooling tower fan speed. %Fan speed directly affects heat removal, and, in turn, impacts the chiller's workload. 
By optimizing the fan speed, we reduce energy use while still maintaining sufficient cooling to meet loads. To implement real-time optimization, we employ an extremum seeking control (ESC) algorithm, which continuously adjusts the system variables to find the optimal operating point for minimal energy usage. 

The main goal of this paper is to demonstrate extremum-seeking control for cooling towers using a simulation-based testbed. However, the work reported here is part of a broader effort to deploy the same type of control strategy on real-world systems in partnership with the US Army Reserve. Our real-world deployment efforts have shown that many chilled water systems do not have direct power measurements available and we have therefore had to adapt the simulation-based strategy to incorporate additional {\em virtual power meters} (VPMs). These VPMs use measurements, such as fan speeds, temperatures, and flow rates, to estimate power use in real-time. The use of VPMs for real-world deployments enables the approach to be fully software-based without having to incur any additional capital costs. We include a description of the VPMs in this paper as these also make use of simulation-based models albeit incorporated in the controller rather than in the testbed. 
Note that the VPMs are not used in the simulation tests since the models are equivalent. However, to account for potential inaccuracies when deploying VPMs in real-world systems, their performance is validated using data from buildings having hardware-based power meters.
 
%essentially the same as the ones in the simulation, meaning that measuring power directly from the simulation would be equivalent to calculating the power in the VPMs from other simulation signals. Deploying the VPMs on real world systems does introduce some inaccuracy though, and this paper ascertains this by validating with data from buildings equipped with hardware-based power meters.

The contributions of this paper are:
\begin{itemize}
    \item Development of a real-time optimizer for chiller plant system using ESC algorithm, validated through simulation to demonstrate its effectiveness in reducing energy consumption.
    \item Design of a VPM for power estimation, validated using real data.
\end{itemize}

The paper is organized as follows: Section 2 reviews relevant literature, Section 3 discusses chilled water plant optimization, including the algorithm and cost function, Section 4 describes the VPM, Section 5 presents the results and discussion, and Section 6 concludes the study.

% \subsection{Paper Organization}
% The paper is organized as follows: Section~\ref{sec:lit} reviews relevant literature, Section~\ref{sec:rto} discusses chiller water plant optimization, inclduing the algorithm and cost function, Section~\ref{sec:vpm} focuses on the development of the virtual power meter, Section~\ref{sec:results} presents the results and discussion, and Section~\ref{sec:conc} concludes the study.

\section{Literature review}\label{sec:lit}
This section provides a literature review of control strategies for chilled water systems and energy monitoring techniques relevant to the design of our VPM-based control strategy.

\subsection{Control Strategies for Chilled Water System}
% \textcolor{red}{(1) what's conventional control method? - limitation? (2) setpoint optimization or fan speed control (3) algorithm used for real time optimization (e.g., mpc, reinforcement learning) (4) challenges and limitation from existing work (e.g., computation, data, etc. )}

Conventional control strategies primarily rely on feedback mechanisms such as proportional-integral-derivative (PID) controllers to regulate chilled water temperature, modulate cooling tower fan speed, and adjust pump speeds for maintaining system performance. While effective for maintaining system performance, these methods are not always energy-efficient because they do not take energy consumption directly into account.

To improve energy efficiency, various strategies have been explored. One approach is adjusting condenser water setpoints based on ambient wet-bulb temperature. On hot days, the system raises the setpoint to increase heat rejection efficiency, while on cooler days, it lowers the setpoint to reduce energy use. Ostendorp et al. \cite{ostendorp2010chilled} implemented this by modulating the cooling tower fans to keep the condenser temperature close to a setpoint. The key idea behind this approach is to dynamically adjust the condenser water temperature depending on external conditions. Lowering the condenser water temperature during cooler periods enhances heat transfer, while raising it during warmer periods reduces energy consumption.

The energy-saving potential of this strategy has been confirmed by several researchers. Grahovac et al. \cite{milica2022} compared a base control strategy with an alternative control sequence from the American Society of Heating, Refrigerating and Air-Conditioning Engineers (ASHRAE) Guideline 36-- High-Performance Sequences of Operation for HVAC Systems for a closed-loop model of a cooling plant \cite{ASHRAE36}. The results showed a 25\% reduction in energy use. Stout and Leach \cite{stout2002cooling} evaluated different fan speeds (single-speed, two-speed, and variable-speed) and found that the two-speed fans provided higher energy savings in areas with a relatively stable wet-bulb temperature. Wang et al. \cite{Wang} demonstrated that fan speed optimization is highly sensitive to chiller load, and power consumption is less sensitive near the optimal fan speed.

Further studies have focused on optimizing other aspects of cooling tower operations. Yu and Chan \cite{yu2008optimization} emphasized the potential benefits of load-based speed control for cooling tower fans and pumps, which led to a 5.3\% reduction in annual electricity use. Cutillas et al. \cite{garcia2017optimum} also highlighted the importance of optimizing control strategies for both energy and water conservation in cooling systems.

Alongside these strategies, optimization algorithms have been explored. Ahn et al. \cite{Ahn2001} developed an optimal control method that minimizes energy consumption for cooling plants by using a quadratic regression model to predict total system power.
% MPC uses mathematical models to predict future system behavior and optimize control actions. 
Braun et al. \cite{Braun2007} proposed an optimization algorithm to control cooling tower fan settings in response to the load on hybrid chillers, reducing operational costs. Liao et al. \cite{Liao2019} demonstrated that optimized temperature reset strategies could save 15.6\% of energy consumption by modeling the cooling tower performance and implementing field measurements. Faulkner et al. \cite{Faulkner2024} explored a chiller plant model with water-side economizers, identifying parameters (e.g., differential temperature, chiller staging) to optimize energy savings.
Huang et al. \cite{Huang} implemented MPC to optimize condenser water setpoints and achieved up to 10\% energy savings. While MPC has proven effective, it requires sophisticated models which may also need retuning over time, making it difficult and costly to implement in real-world applications.

ESC, on the other hand, uses real-time responses to adjust control setpoints without the need for complex models. Zhao et al. \cite{Zhao2022} demonstrated ESC's potential to save energy by 14\% in chilled-water plants by adjusting control variables such as condenser water mass flow rate, chilled water temperature setpoint, and tower fan air mass flow rate. 
Despite its potential, the traditional ESC still requires multiple parameters, which makes it cumbersome in practice.

Control vendors commonly employ strategies to vary the speed of chilled water pumps and cooling tower fans to maintain desired condenser temperatures \cite{jsc}. More advanced solutions, including analytics that use artificial intelligence, are beginning to emerge for predictive adjustments, but they still adjust chilled water flow and temperature dynamically based on real-time demand, typically using sensor feedback. 

Despite significant progress in optimizing energy efficiency, existing strategies have two key limitations. First, they often do not incorporate direct power consumption measurements into the control loop, instead relying on predicted or surrogate metrics. Second, they rely heavily on complex models or algorithms that require substantial tuning, computational resources, or expert configuration. These challenges highlight the need for a control approach that is both model-free and easy to implement, capable of reducing energy consumption in real time without requiring detailed system modeling or continual retuning.

\subsection{Energy Monitoring Techniques} % virtual power metering for energy estimation and control
%(1) role of building management systems (2) sensors and IoT applications in chilled water plants (3) virtual power meter? how they estimate power consumption without physical metering  - mathematical or machine learning-based approaches - how VPMs are implemented within the system}

Energy system monitoring techniques have been widely used for fault detection and system optimization, as highlighted by Aguilar et al. \cite{aguilar2020autonomic}. Machine learning (ML) algorithms have also been used to detect operational anomalies and system faults \cite{miller2018review} \cite{chen2023review}. However, direct energy monitoring is challenging unless a physical power meter is available. To estimate energy consumption, several methods can be employed.

Building energy management systems (BEMSs) can be used to monitor and optimize the operation of chillers, pumps, and cooling towers based on input from multiple sensors and system variables. These systems often incorporate data-driven methods, such as ML models, and when combined with real-time monitoring systems, can be used to estimate energy use. In these systems, sensor data plays a crucial role in enabling energy estimation and prediction~\cite{gunay2019data}. 

On the other hand, O'Neil et al. \cite{oneil2011real} and Deblois et al. \cite{deblois2015} applied simulation models, such as EnergyPlus and neural networks, to estimate building performance and energy consumption. Although their proposed models effectively predicted energy use, their focus has generally been on scenario analysis, rather than providing real-time real-world energy estimates for control.

Even though energy monitoring techniques have shown promise for fault detection and system optimization, the challenge remains to develop accurate, real-time energy estimation methods, particularly in the absence of physical power meters, to enable effective control and optimization in real-world systems. %While the overall energy consumption might be monitored, 
Understanding the energy consumption of individual components of the building energy system can provide valuable insights for optimizing the operation.

\begin{figure*}[ht]
\centering
\includegraphics[width=0.8\textwidth]{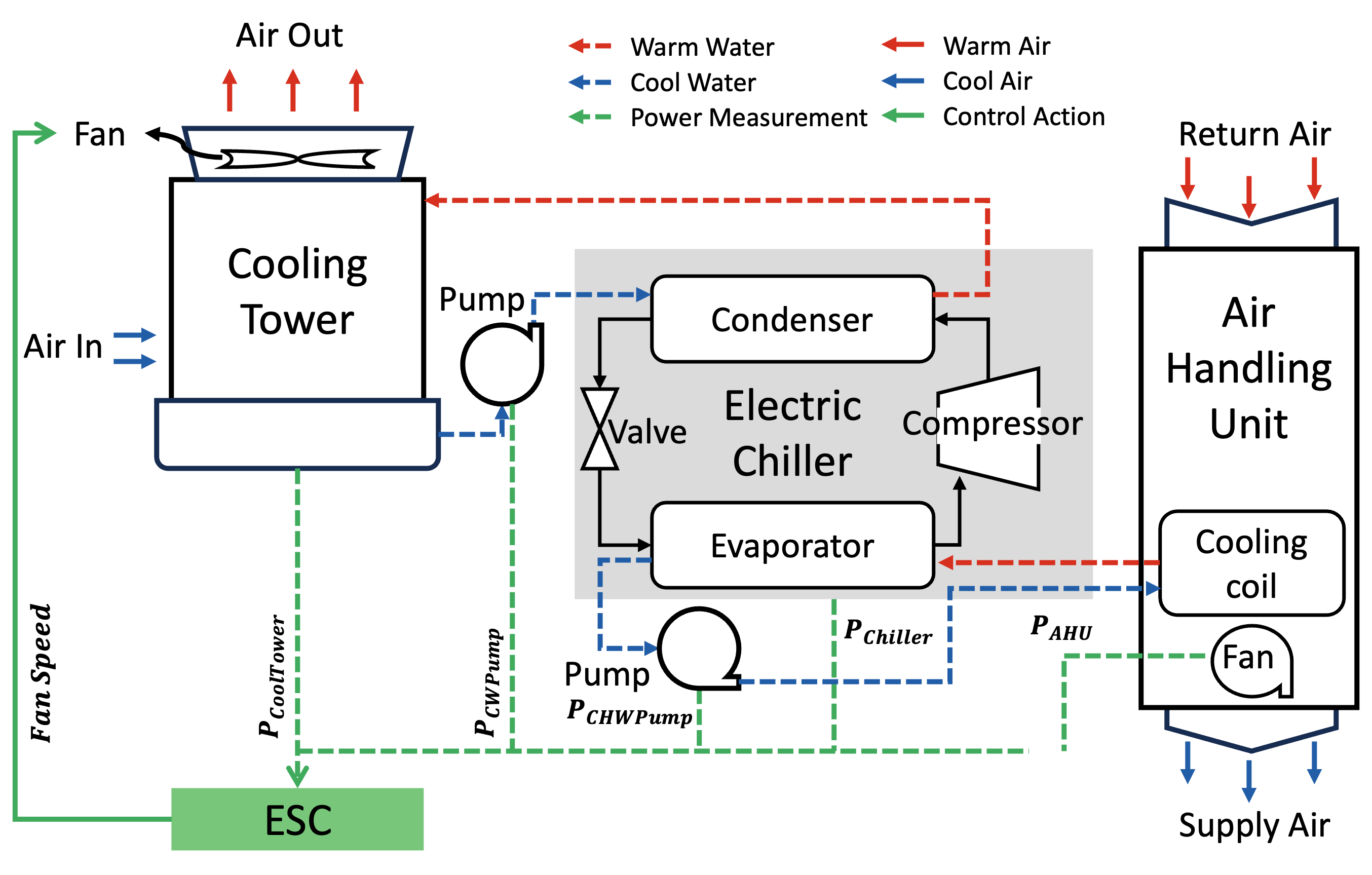}
\caption{Chiller plant system with the ESC to control cooling tower fan.}
\label{fig:systemmodel}
\end{figure*}

\subsection{Research goals and scope}
% \textcolor{red}{why optimization is crucial for chiller water plant, challenges for applying real time optimization and energy estimation to real world systems including lack of power meters }

Based on the gaps identified in the previous sections, the research goal of this work is to develop a simple-to-configure controller for cooling towers to reduce overall chiller plant energy usage without being able to directly measure power use. This paper focuses on optimizing the operation of cooling tower fans, using a relay-based extremum-seeking controller to enhance energy efficiency under varying environmental and load conditions. By addressing the challenge of having no physical power meters, the research will estimate the power instead from typically available measurements.

\section{Chiller Water Plant Optimization} \label{sec:rto}
This section outlines the research problem and optimization algorithm, including the cost function and the configuration of the parameters required for the algorithm. Additionally, we describe the simulation setup used to demonstrate the performance of the chilled water plant optimization.

\subsection{Problem Statement}
Figure \ref{fig:systemmodel} illustrates the proposed approach, which uses an ESC-based real-time optimizer to control the chiller plant system. The controller is connected to subsystems such as the cooling tower, condenser water pump, chiller, chilled water pump, and air handling unit (AHU) fan. As discussed in Section 1, optimizing energy consumption in a chiller plant system involves balancing the energy usage of these components. The water leaving the chiller condenser is cooled through a cooling tower, which exchanges heat between the condenser water and the ambient air. 

The cooling tower fans circulate ambient air over the sprayed water to facilitate heat transfer and lower the water temperature. As cooling towers remove heat from the condenser loop, the chiller's workload decreases, but this creates a trade-off between the energy consumption of the condenser loop and the chiller system. Increasing the cooling tower fan speed improves heat rejection and reduces the chiller's load, but also raises the cooling tower fan energy consumption. The challenge is to find the optimal balance between these competing energy demands. 

\subsection{Optimization Algorithm}
\begin{figure}[!t]
    \centering
    \includegraphics[width=0.48\textwidth]{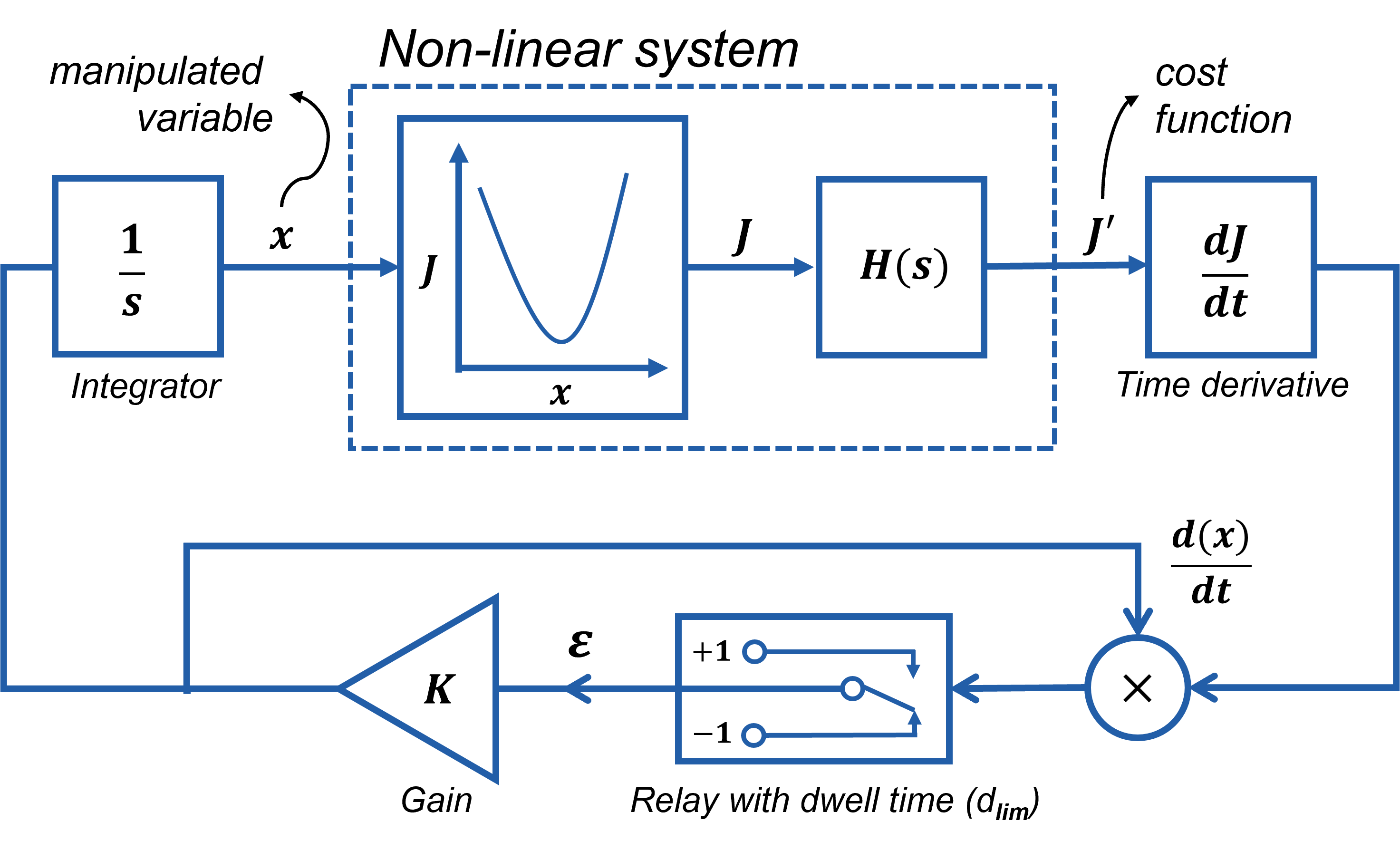}
    % \vspace{-0.15 cm}
    \caption{Block diagram of real-time optimizer.}
    \label{fig:esc_block}
    % \vspace{-0.65cm}
\end{figure}

ESC is a feedback-based control mechanism that follows the ``perturb and observe" principle. It introduces a perturbation to the system, observes its response, and determines the appropriate control action based on that response. The specific type of ESC implemented in our work is a single-input, single-output (SISO) relay-based algorithm, as described in Figure~\ref{fig:esc_block} (See \cite{salsbury2023} for more details).

In this block diagram, a relay block generates a signal with a value of either -1 or +1, denoted as $\epsilon$. This signal is determined by the sign of the gradient of the cost function $J'$ and follows the trajectory of the negative gradient. To account for system response and avoid rapid switching, a time-scale separation technique is employed, using a dwell time ($d_{lim}$) that causes the relay block to hold its current state for a minimum time. The dwell time is calculated as: 
\begin{equation}
d_{lim} = \tau + \tau_f
\end{equation}
\noindent where $\tau$ is the system time constant and $\tau_f$ is a smoothing filter time constant.

The relay output $\epsilon$ is then multiplied by a gain factor $K$ and integrated to produce the system's manipulated variable, $x$. The manipulated variable is constrained within specified minimum and maximum bounds. The gain factor $K$ is determined as:
\begin{equation}
K = \frac{\Delta t (x_{max} - x_{min})}{5(\tau + \tau_f)}
\end{equation}
\noindent where $\Delta t$ is the discrete time step.

All internal variables (e.g., $K$, $d_{lim}$) are determined from a single parameter: the time constant $\tau$ of the system being optimized, which we will discuss in the following subsection. In this work, the manipulated variable $x$ represents the cooling tower fan speed.

% \begin{algorithm}[ht]
% \caption{Pseudo code implementation of the ESC-based real-time optimization algorithm}
% \label{code:alg}
% \begin{algorithmic}[1]
% \State $\epsilon_t = 1$
% \State $g_t = 0$
% \State $d_t = 0$
% \State $a = \exp(-\Delta t/\tau_f)$
% \State $d_{lim} = \tau + \tau_f$
% \State $K = \Delta t(x_{max}-x_{min})/ (5 (\tau + \tau_f))$
% \While{$t > 0$}
%   \State $g_t = a*g_{t-1} + (J'_t - J'_{t-1})$
%   \If{$g_t > 0$ \textbf{AND} $d_t \geq d_{lim}$}
%     \State $\epsilon_t = -\epsilon_{t-1}$
%     \State $d_t = 0$
%   \EndIf
%   \State $x_t = x_{t-1} + \epsilon_t * K$
%   \State $x_t = \max(x_t, x_{min})$
%   \State $x_t = \min(x_t, x_{max})$
%   \State $d_t = d_{t-1} + \Delta t$
% \EndWhile
% \end{algorithmic}
% \end{algorithm}

\subsection{Cost function}
To minimize energy usage across the chiller plant system, the cost function $J$ is defined as the sum of the power consumption of the system's key components: the cooling tower $P_{CoolTower}$, condenser water pump $P_{CWPump}$, chiller $P_{Chiller}$, chilled water pump $P_{CHWPump}$, and AHU fan $P_{AHU}$. This approach directly targets energy consumption by minimizing the combined power usage of these interconnected subsystems. Since the ESC is placed on top of all the existing controllers, it does not interfere with the lower-level feedback controllers, ensuring easier implementation.
\begin{equation}
J = P_{CWPump} +P_{CHWPump} + P_{Chiller}
+ P_{CoolTower} + P_{AHU}
\end{equation}

\begin{figure*}[ht!]
    \centering
    \includegraphics[width=0.95\textwidth]{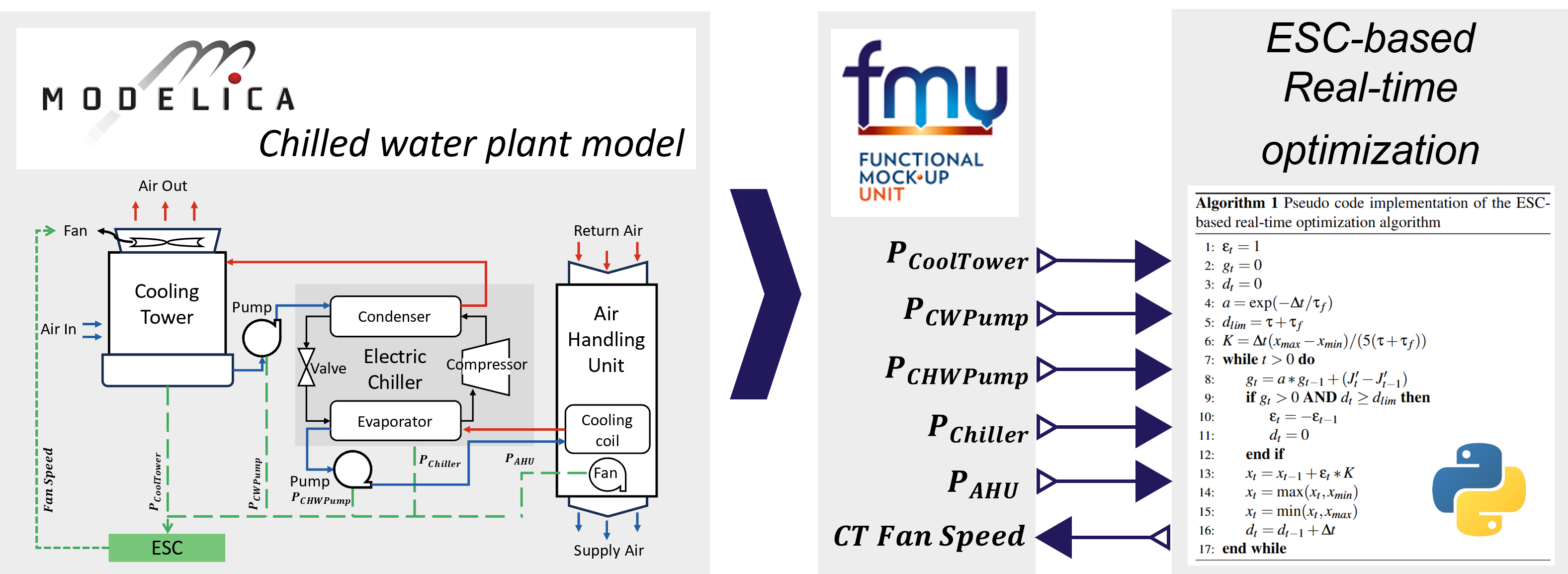}
    \caption{Co-simulation workflow.}
    \label{fig:workflow}
\end{figure*}

\subsection{Simulation Setup}

To demonstrate the performance of the chilled water plant optimization before real-world implementation, a simulation demonstration is necessary. In this work, we developed an ESC controller using Python (version 3.11) and tested it with a chiller plant model from the Modelica Buildings Library (version 10.0) \cite{Modelica}.
We modified the model to adjust the cooling tower fan speed based on inputs from the ESC controller. Specifically, we: 1) enabled cooling tower fan speed control via an input signal, 2) disabled hysteresis-based cycling control tied to loop temperature, and 3) deactivated the water-side economizer.
The power consumption of all components (cooling tower, chiller, condenser water pump, chilled water pump, and AHU fan) is extracted as output and fed into the ESC controller as the cost function. 

In our simulation, the cooling tower operates with a variable-speed fan and the chiller plant uses variable-speed pumps to circulate chilled water, with the chiller model always enabled. The chiller's outlet temperature setpoint regulates its performance. Additionally, the pump receives an input for the static differential pressure setpoint and adjusts its speed via an internal loop. Both the differential pressure and chiller outlet temperature setpoints are adjusted by PI controller. The AHU fan is variable-speed and controlled by a PI loop to maintain a constant zone temperature of 23$^{\circ}$C. The thermal zone is modeled as a data center, based on the EnergyPlus model `1ZoneDataCenterCRAC\_wApproachTemp.idf', where IT equipment is the primary heat source.
The simulations employed the CVODE solver with the convergence tolerance of 1e-6 and a fixed simulation time step of 60 seconds, chosen to ensure the balance between computational efficiency and result fidelity. 

The Modelica model was then exported as a Functional Mockup Unit (FMU) and coupled with the Python program for the ESC controller using the FMPy Python library \cite{FMPy}. Figure~\ref{fig:workflow} illustrates the co-simulation setup, showing the interaction between the simulation model and the ESC controller.
The co-simulation workflow enables a dynamic exchange of data between the FMU and the ESC controller. The FMU provides crucial input data, such as the power consumption of the system components for the ESC algorithm to calculate the optimal fan speed setpoint. Once determined, the ESC outputs the fan speed setpoint, which is then fed back into the FMU to adjust the cooling tower fan speed. This process repeats in an ongoing cycle, ensuring continuous interaction between the FMU and ESC, and allowing the ESC to optimize performance throughout the simulation.

\subsection{ESC configuration}
\begin{figure}[!ht]
    \centering
    \includegraphics[width=3.1 in]{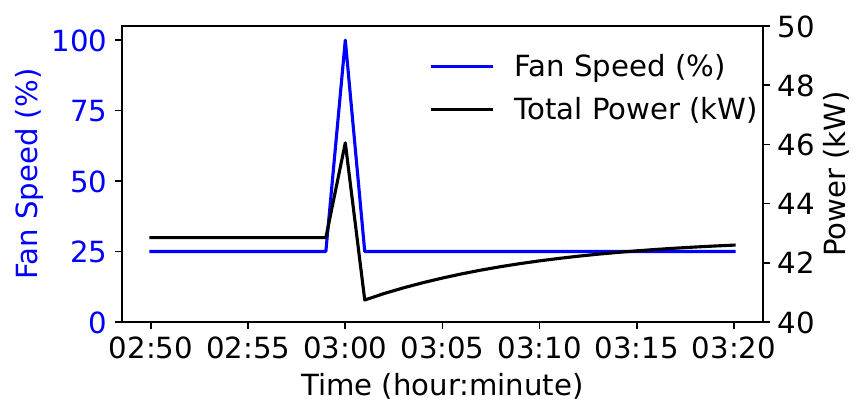}
    \caption{Impulse test to estimate the time constant of the system.)}
    \label{fig:bump}
\end{figure}

Implementing the ESC algorithm requires just one system parameter, the time constant $\tau$. To estimate this time constant, the system's dynamic response needs to be tested and, in this work, an impulse response test was used to estimate the time constant. The impulse test is a method where the system's input is significantly changed from one value to another and held for a short period before returning back to the starting value. Figure~\ref{fig:bump} shows a sample result of this impulse test. By changing the fan speed from 25\% to 100\% at 03:00, we monitored the changes in the cost function (e.g., total power) and estimated the time constant based on the response. The system showed a time constant of 183 seconds.
For a cooling tower fan, a time constant in the range of a few minutes (e.g., 3-5 minutes) is typical. This relatively quick response time is also beneficial for the ESC algorithm as it allows for faster optimization of energy use. %This faster response allows the system to adjust more dynamically, improving overall energy efficiency while maintaining stability.

\section{Virtual Power Meter Development} \label{sec:vpm}
\subsection{Overview}
\begin{figure}[ht!]
    \centering
    \includegraphics[width=\columnwidth]{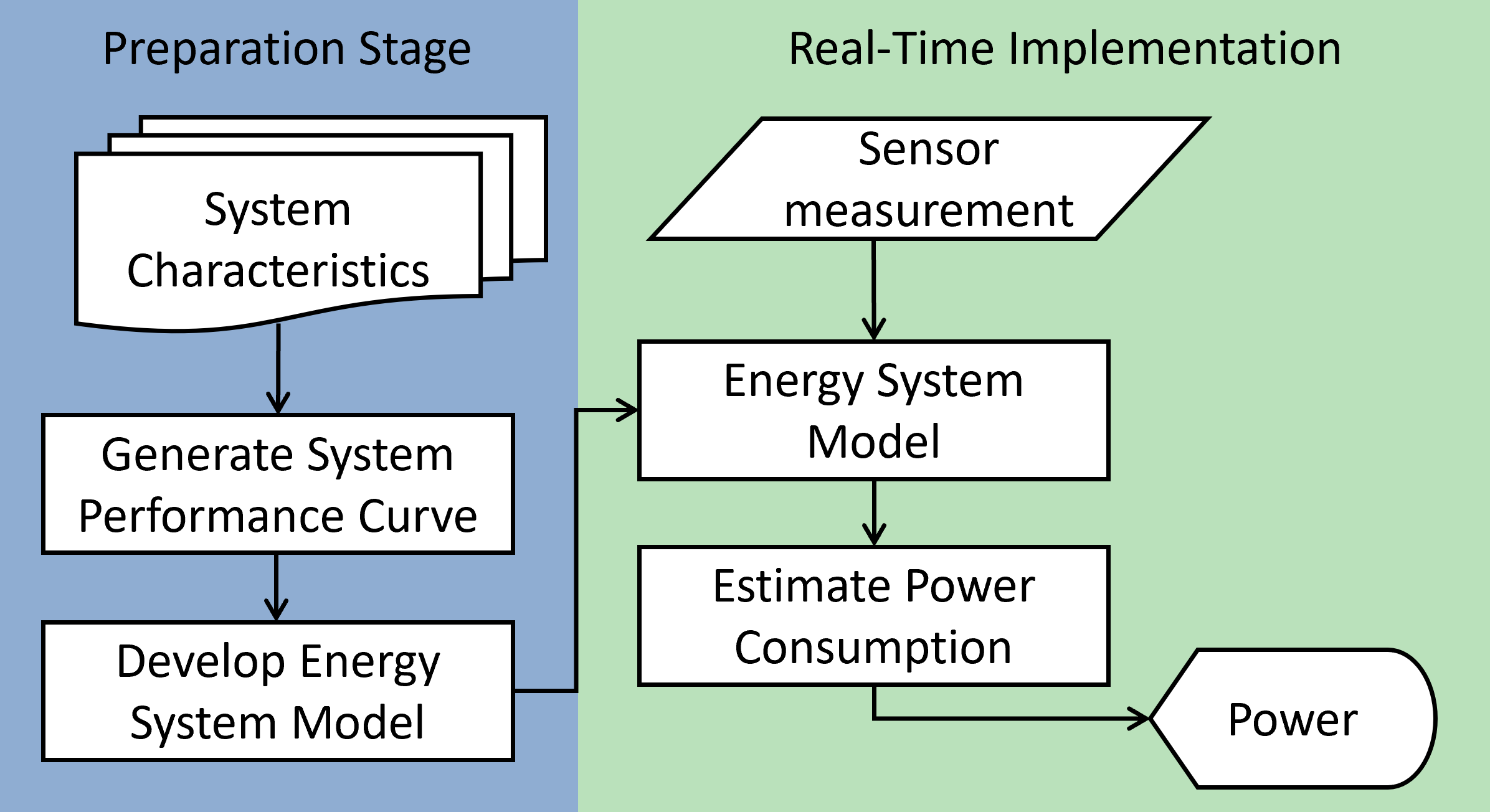}
    \caption{Overview of virtual power meter.}
    \label{fig:vpm}
\end{figure}

Power consumption of the chilled water system components is needed to create the cost function for the proposed real-time optimizer. Because physical power meters are not always available, this paper proposes the use of a VPM as an alternative solution. This section outlines the development of a VPM for chiller and cooling tower fan, as shown in Figure~\ref{fig:vpm}. %Note that the power consumption of the air-side fan and pumps operate at a constant speed, and 
We observed an inverse relationship between the power consumption of the chiller and the cooling tower. Thus, this paper focuses on the development and validation of the VPM for the chiller system and cooling tower fan.

First, we develop a physics-based model to estimate the power consumption of the energy system. Every system has unique system characteristics that can vary based on several factors such as capacity, efficiency, and more. This information can typically be sourced from manufacturer specifications or technical documents. Then, the energy system model is configured to match the specific characteristics of the system. Once the model is properly tailored, the VPM can be implemented using real-time sensor measurements to provide accurate and dynamic power consumption estimates.

We note that while both VPM and Modelica models use similar physics-based equations, they differ in purpose, data sources and model behavior. Modelica models are typically dynamic, used for forward simulation in virtual environments to study system behavior over time under varying conditions. In contrast, VPM models are generally static and estimate real-time energy use in actual buildings using BAS sensor data. They serve as virtual replacements for physical power meters.

\subsection{Chiller Plant Energy System Model}
\subsubsection{Electric chiller}
We developed the electric chiller model based on the {\em chiller model} in the DOE-2.1 building energy simulation program using the functions of temperature and part load ratio (PLR) \cite{EnergyPlusEngRef}.
The power consumption of the chiller $P_\text{chiller}$ is calculated for the chiller compressor, as in \eqref{chiller1}--\eqref{chiller6}. 
\begin{align}
&\psi_{1}(T_\text{evap-l},T_\text{cond-e}) = a_0 + a_1 T_\text{evap-l} + a_2 T_\text{evap-l}^2 + a_3 T_\text{cond-e} \nonumber\\
&\quad \quad + a_4 T_\text{cond-e}^2  + a_5 T_\text{evap-l} T_\text{cond-e} \label{chiller1} \\
&Q_\text{chiller-a} = C_\text{chiller} \psi_{1} \label{chiller2} \\
&q_\text{load} = m_\text{cw} c_\text{cw} (T_\text{evap-e}- T_\text{evap-l}) \label{chiller3} \\
&PLR = q_\text{load} / Q_\text{chiller-a} \label{chiller4}\\
&\psi_{2}(T_\text{evap-l},T_\text{cond-e}) = b_0 + b_1 T_\text{evap-l} + b_2 T_\text{evap-l}^2 + b_3 T_\text{cond-e}  \nonumber \\
&\quad \quad  + b_4 T_\text{cond-e}^2 + b_5 T_\text{evap-l} T_\text{cond-e} \label{chiller5} \\
&\psi_{3}(PLR) = c_0 + c_1 PLR + c_2 PLR^{2} \label{chiller6}\\
&P_\text{chiller} = \frac{Q_\text{chiller-a} \psi_2 \psi_3}{COP_\text{ref}}  \label{chiller7}
\end{align}

% \begin{align}
% P_\text{chiller} &= \frac{Q_\text{chiller-a} \psi_2 \psi_3}{COP_\text{ref}}\\
% &= \frac{q_\text{load} \psi_2 \psi_3}{PLR \cdot COP_\text{ref}}\\
% &= \frac{q_\text{load}}{\frac{PLR}{\psi_2 \psi_3} COP_\text{ref}}\\
% \end{align}
\noindent where $\psi_{1}$ is defined to estimate the chiller's available capacity $Q_\text{chiller-a}$ with the function of chilled water supply setpoint temperature $T_\text{evap-l}$ and condenser entering temperature $T_\text{cond-e}$, PLR is the ratio of the load $q_\text{load}$ to its available capacity $Q_\text{chiller-a}$, $\psi_{2}$ is defined to adjust the ratio of energy to actual load with the function of temperature $(T_\text{cw-s},T_\text{cond-e})$, $\psi_{3}$ is to adjust the ratio of energy input to actual load with the function of PLR, $P_\text{chiller}$ is estimated with the ratio of energy input to actual load, chiller's available capacity and reference coefficient of performance ($COP_\text{ref}$).

To account for the specific design characteristics of the chiller, key parameters of the systems, including capacity, efficiency, compressor type, condenser type, and compressor speed, are required. Utilizing tools like `Copper' \cite{Copper}, the coefficients ($a_{0-5}, b_{0-5}, c_{0-2}$) for three essential performance curves for the chiller can be fitted in Eqs.~\eqref{chiller1}, \eqref{chiller5}, \eqref{chiller6}. In real-time implementation stage, real-time sensor measurement data (e.g., chilled water flow rate, evaporator entering temperature, evaporator leaving temperature, condenser entering temperature) are collected. 
Then, based on the reference capacity and temperature-based performance curve \eqref{chiller1}, the available capacity can be calculated as in \eqref{chiller2}. Next, the load assigned to the chiller to cool down can be estimated based on the operating conditions as in \eqref{chiller3}. The part load ratio of chiller system can be determined accordingly as in \eqref{chiller4}. Finally, the chiller's power considering the adjusted cooling capacity, cooling load, and compressor workload required for the specific cooling based on the operating temperature and part load ratio is calculated to generate the estimated (virtual) power.

\subsubsection{Cooling Tower Fan}
The cooling tower fan power is directly proportional to the cube of the fan speed. Motor power is specified in horsepower (HP) from manufacturer's data and converted to kilowatts (kW) using a conversion factor of 0.7457. The fan power, adjusted based on the cube of the fan speed percentage (\%), is then calculated as follows:
\begin{align}
    P_{fan} = P_{HP} \cdot 0.7457 \cdot \left(\frac{S}{100}\right)^3
\end{align}
\noindent where $P_{fan}$ is the fan power in kW, $S$ is the fan speed as a percentage of the maximum speed.

\subsection{Evaluation Metrics}
To evaluate the functionality of the VPM model, we used the following metrics:

{\em Coefficient of determination:}
\begin{align}
R^2 &= 1 - \frac{\sum\limits_{i=1}^{n} (y_i - \hat{y}_i)^2}{\sum\limits_{i=1}^{n} (y_i - \bar{y})^2} \label{eq:r2}
\end{align}

{\em Root mean square error:}
\begin{align}
RMSE &= \sqrt{\frac{1}{n} \sum_{i=1}^{n} (y_i - \hat{y}_i)^2} \label{eq:rmse}
\end{align}

{\em Normalized Root mean square error:}
\begin{align}
NRMSE &= \frac{RMSE}{\text{Range}(y)} = \frac{\sqrt{\frac{1}{n} \sum\limits_{i=1}^{n} (y_i - \hat{y}_i)^2}}{\left( y_{\text{max}} - y_{\text{min}}\right)} \label{eq:nrmse}
\end{align}

\noindent where $y$ is the measured value, $\hat{y}$ is the estimated value from the VPM, $\bar{y}$ is the mean of the measured value, $n$ is the number of observations, $y_{\text{max}}$ and $y_{\text{min}}$ are the maximum and minimum value of the measured data.

Equation~\eqref{eq:r2} quantifies how well the estimated values from VPM model match the actual data. The higher the $R^2$, the better the model fits the actual data where $0 \leq R^2 \leq 1$. Equation~\eqref{eq:rmse} provides a measure of average differences between the estimated and actual values. Equation~\eqref{eq:nrmse} is normalized based on the range of the actual data.

\section{Results and Discussion} \label{sec:results}
\subsection{Simulation Results for Chiller Plant Optimization}

\begin{figure}[!t]
\centering
\begin{subfigure}[b]{0.45\textwidth}
\includegraphics[width=1\linewidth]{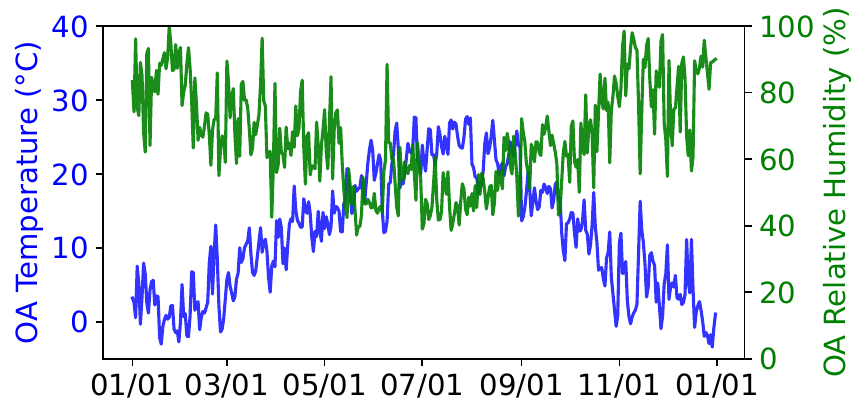}
   \caption{Pasco, WA (5B).}
   \label{fig:pasco_weather} 
\end{subfigure}
\begin{subfigure}[b]{0.45\textwidth}
\includegraphics[width=1\linewidth]{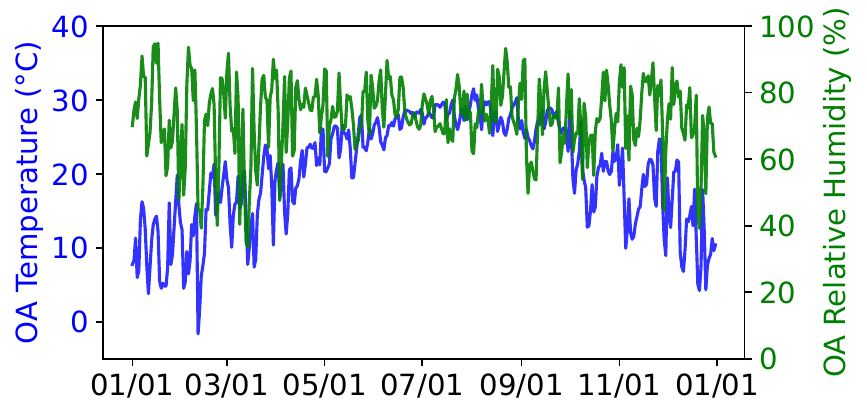}
   \caption{Houston, TX (2A).}
   \label{fig:houston_weather}
\end{subfigure}
\caption{Weather profiles for two climates.}
\label{fig:weather_profiles}
\end{figure}

For the simulation study, we explored two geographic regions (a) Pasco, WA and (b) Houston, TX. Figure~\ref{fig:weather_profiles} illustrates the annual weather profile for each location, with the daily average outdoor air (OA) temperature and relative humidity. The weather data is based on the TMY3 (Typical Meteorological Year 3) dataset.

The weather in Pasco, WA shows significant seasonal variation. 
Winters are cold with higher humidity levels reaching up to ~90-100\% due to precipitation and colder air. Summers, in contrast, are hot and dry with humidity levels dropping to ~30-50\%. 
The weather in Houston, TX shows less seasonal variability, reflecting a humid subtropical climate with warm to hot and high humidity year-round. Relative humidity generally stays above 60\% throughout the year and peaks close to 100\%. 
Since this study focuses on chiller plant operations with ESC, the simulation is conducted during the cooling season. 
%Hot and humid climates, such as Houston, are typically more favorable for achieving energy savings with cooling towers because cooling towers rely on evaporative cooling. In this condition, the combination of warm air and higher relative humidity enhances the heat rejection process, so that the cooling tower can operate efficiently.

\begin{figure}[!t]
\centering
\begin{subfigure}[b]{0.47\textwidth}
\includegraphics[width=1\linewidth]{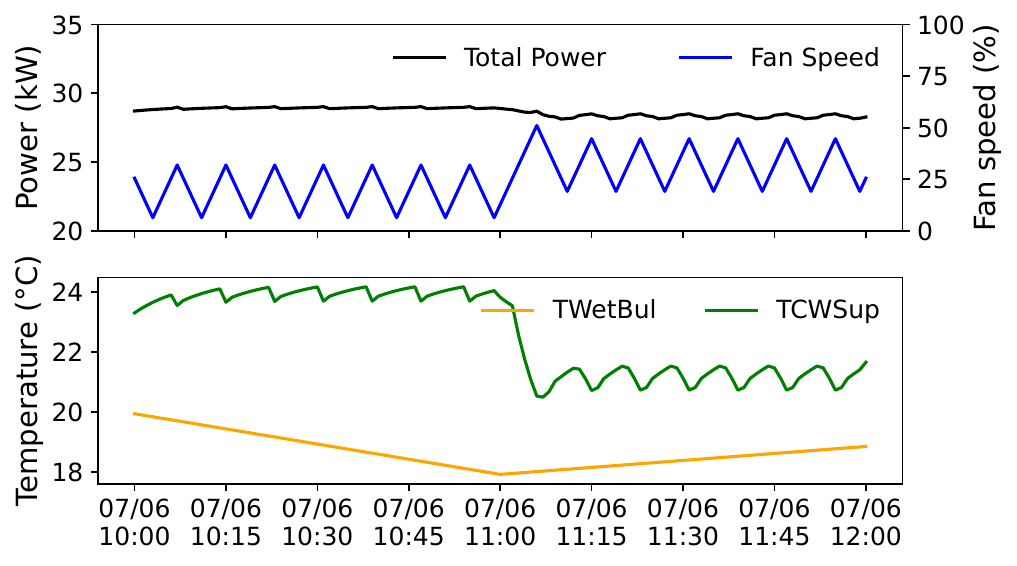}
   \caption{Pasco, WA (5B)}
   \label{fig:pasco_ESC_2hour} 
\end{subfigure}
\begin{subfigure}[b]{0.47\textwidth}
\includegraphics[width=1\linewidth]{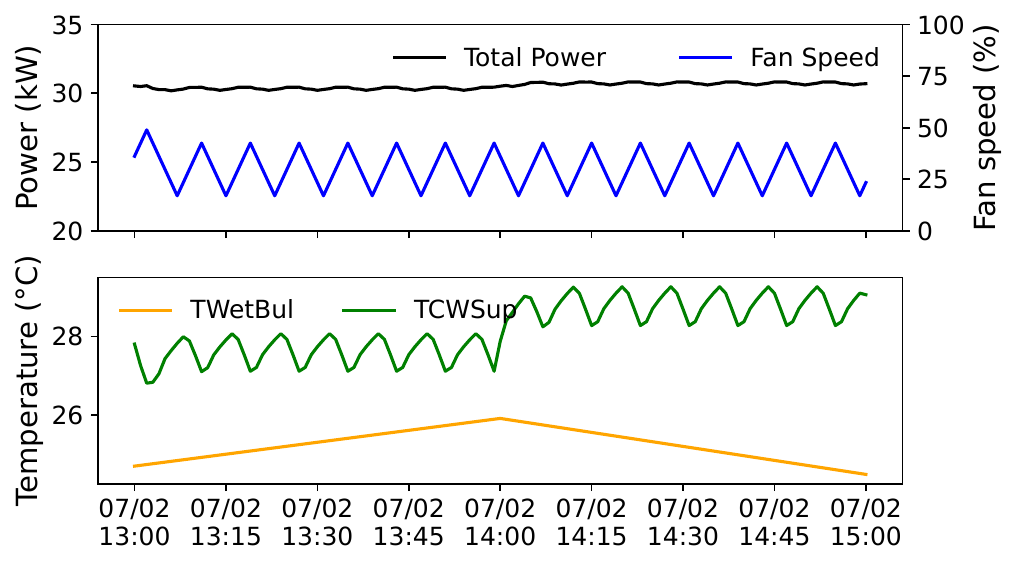}
   \caption{Houston, TX (2A)}
   \label{fig:houston_ESC_2hour}
\end{subfigure}
\caption{Snapshot of simulation results with the Extremum Seeking Control}
\label{fig:ESC_performance_2hour}
\end{figure}

With our simulation model, we tested a one-week period in July for both climate locations. Figure ~\ref{fig:ESC_performance_2hour} presents time series plots illustrating the ESC algorithm-based cooling tower fan speed control results. 
In each climate location, the top plot shows the cooling tower fan speed controlled by the ESC algorithm and the total power consumption (i.e., cost function). The bottom plot shows the outdoor wet bulb temperature and the condenser water supply temperature (TCWSup). 
This figure provides a 2-hour snapshot of the simulation period to examine the detailed dynamics of the result. The shorter time frame is chosen because the fan speed is adjusted every 3-5 minutes due to the ESC algorithm, making it easier to interpret the system behavior.

The first notable feature of this plot is the sawtooth oscillation pattern of the fan speed, which is caused by the relay-based ESC algorithm. 
For the Pasco case (Fig~\ref{fig:pasco_ESC_2hour}), the fan speed fluctuated between 0-20\% during the first hour. At around 11:00, the fan speed increased sharply to 50\%, then settled into oscillations between 20-40\% in the second hour. These adjustments reflect the ESC algorithm searching for the optimal point to minimize energy consumption. Note that when the ESC finds the optimal value, the oscillations persist so that it can always be testing if the optimal location has changed.
Importantly, the increased fan speed during this period enhances the heat rejection process by moving more air across the cooling tower's fill and increasing evaporation. This results in a reduced condenser water supply temperature (TCWSup). Additionally, since the outdoor wet-bulb temperature was relatively lower during the second hour, the cooling tower was able to reject heat more effectively. 
This not only improves the cooling tower's performance but also enhances the efficiency of the chiller. This, in turn, results in a slight reduction in total power consumption. Throughout the simulation testing period for the Pasco case, the ESC algorithm controlled the cooling tower fan speed between a mean value of 30\% and a lower bound near zero.

Figure~\ref{fig:houston_ESC_2hour} presents the results for Houston, where the ESC algorithm maintains the average fan speed more consistently between approximately 20\% and 40\%. Even though the outdoor wet-bulb temperature increased during the first hour and then decreased during the second hour, the fan speed oscillation followed the same pattern, with the TCWSup increasing with a comparable relative change.

\begin{figure}[!t]
\centering
\begin{subfigure}[b]{0.49\linewidth}
\includegraphics[width=\linewidth]{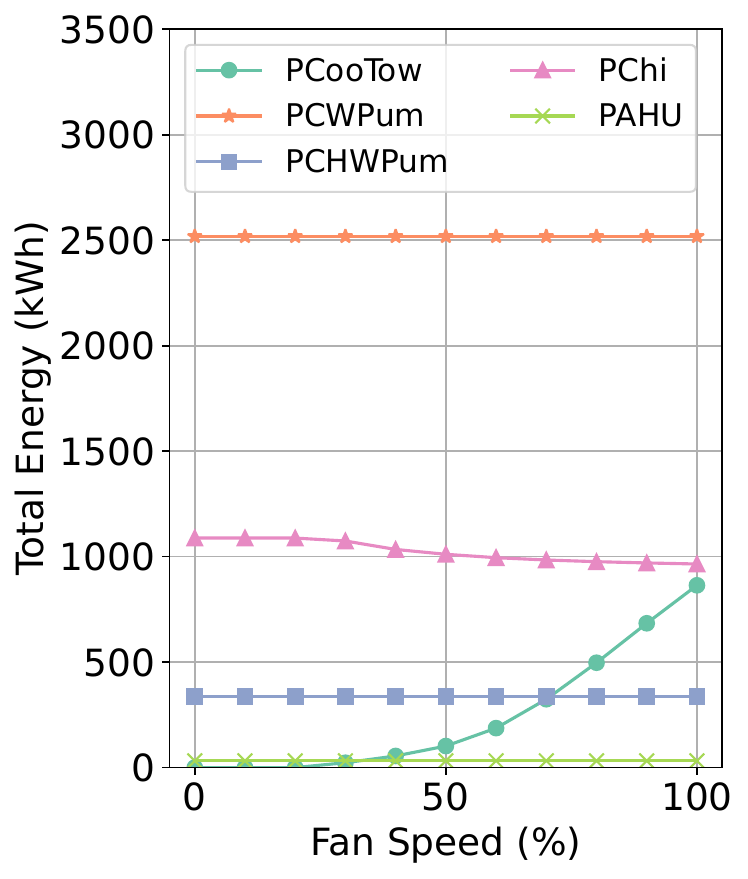}
   \caption{Pasco, WA (5B)}
   \label{fig:pasco_power_comparison} 
\end{subfigure}
\hfill
\begin{subfigure}[b]{0.49\linewidth}
\includegraphics[width=\linewidth]{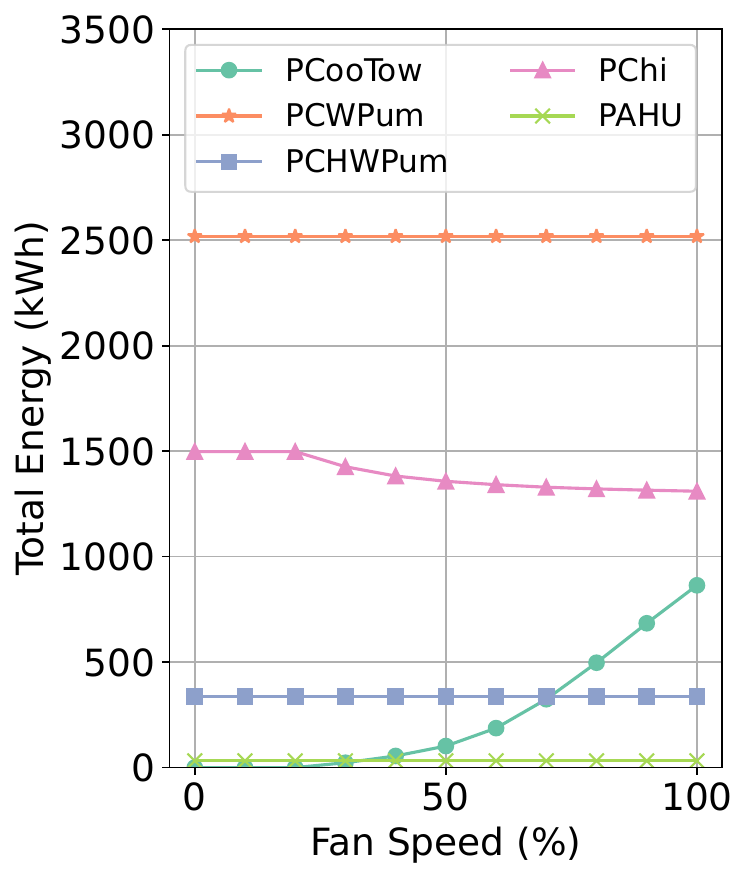}
   \caption{Houston, TX (2A)}
   \label{fig:Houston_power_comparison}
\end{subfigure}
\caption{Power consumption by each subcomponent.}
\label{fig:each_power_comparison}
\end{figure}

\begin{figure}[!t]
\centering
\begin{subfigure}[b]{0.48\linewidth}
\includegraphics[width=\linewidth]{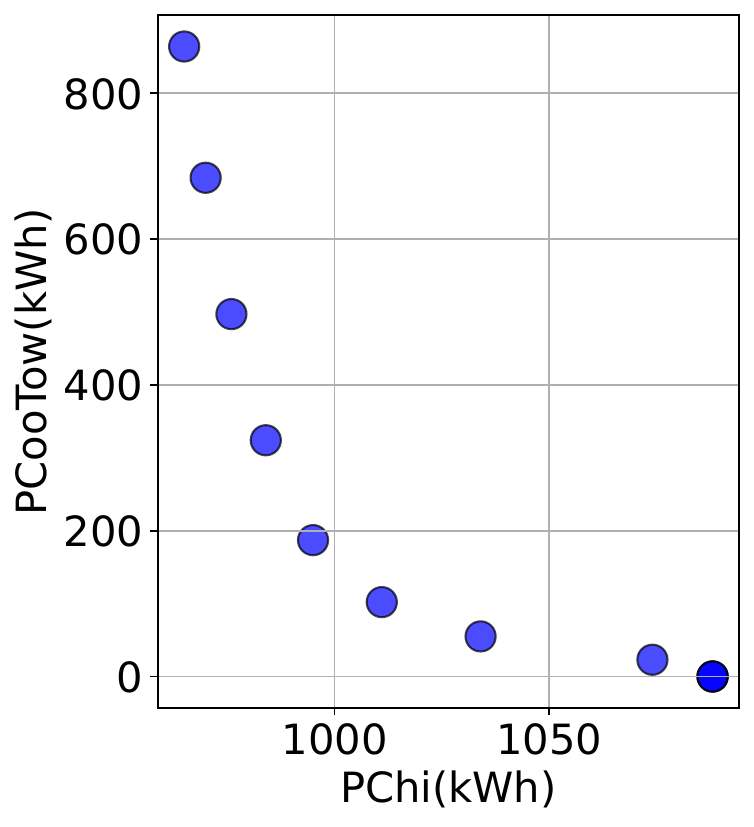}
   \caption{Pasco, WA (5B)}
   \label{fig:pasco_scatter} 
\end{subfigure}
\hfill
\begin{subfigure}[b]{0.49\linewidth}
\includegraphics[width=\linewidth]{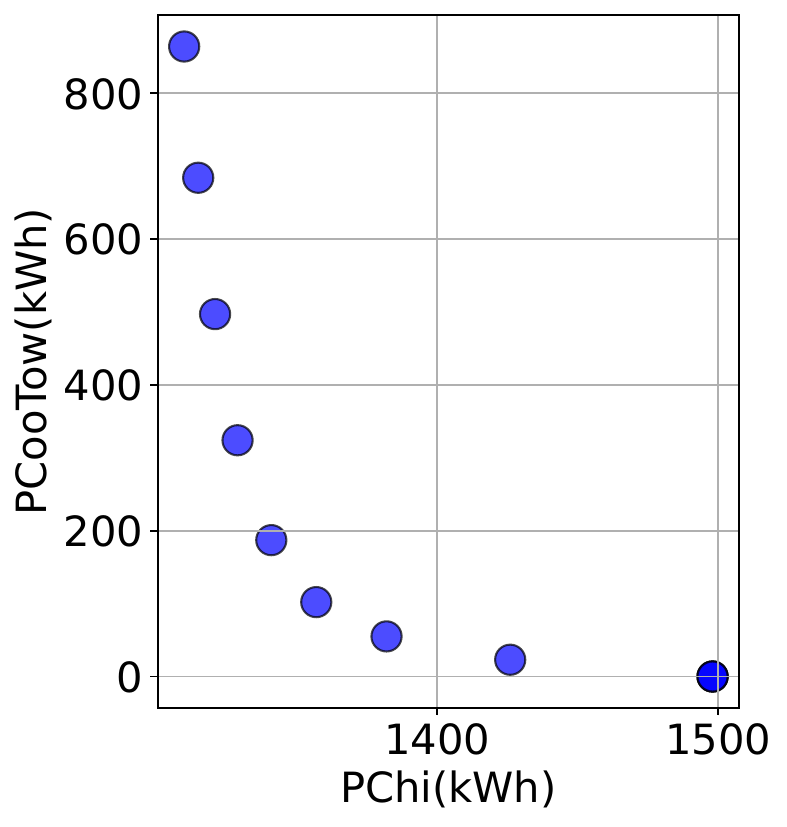}
   \caption{Houston, TX (2A)}
   \label{fig:Houston_scatter}
\end{subfigure}
\caption{Cooling tower power vs. chiller power by varying cooling tower fan speed.}
\label{fig:power_scatter}
\end{figure}

\begin{figure}[!t]
\centering
\begin{subfigure}[b]{0.49\linewidth}
\includegraphics[width=\linewidth]{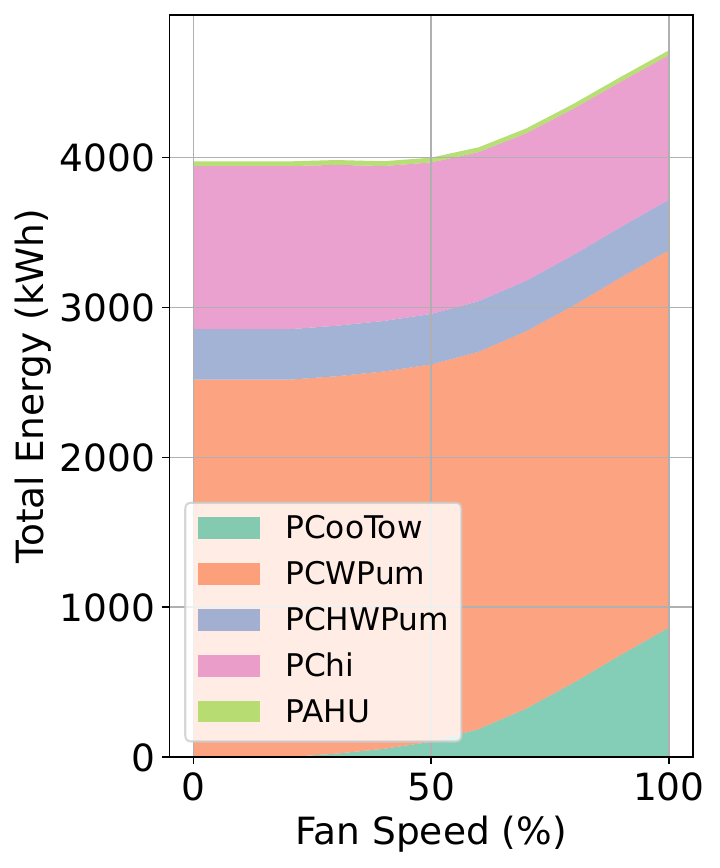}
   \caption{Pasco, WA (5B)}
   \label{fig:pasco_stack} 
\end{subfigure}
\hfill
\begin{subfigure}[b]{0.49\linewidth}
\includegraphics[width=\linewidth]{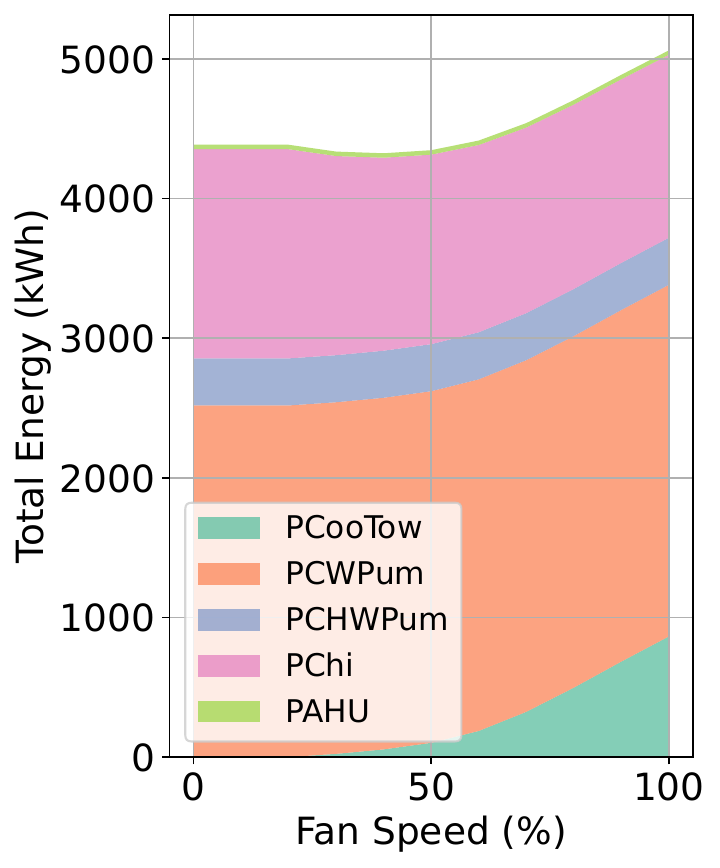}
   \caption{Houston, TX (2A)}
   \label{fig:Houston_stack}
\end{subfigure}
\caption{Stacked power consumption of components.}
\label{fig:power_stack}
\end{figure}

To evaluate whether the ESC algorithm effectively adjusts the cooling tower fan speed to minimize total energy consumption, we analyzed the detailed power consumption of each system component and the cumulative total power by varying fan speeds.
Figure~\ref{fig:each_power_comparison} illustrates the energy consumption of key system components—cooling tower (PCooTow), chiller (PChi), condenser-side water pump (PCWPum), chilled water pump (PCHPum), and air loop fan (PFan)—throughout the simulation testing period for both climate locations. The pumps and air loop fan maintained constant power consumption, while notable differences were observed in the chiller and cooling tower fan power consumption. As shown in Figure~\ref{fig:power_scatter}, there is an inverse relationship between the cooling tower fan and chiller power consumption. For both climates, an increase in fan speed led to higher cooling tower power consumption but lower chiller power consumption.

\begin{figure}[!t]
\centering
\begin{subfigure}[b]{0.45\textwidth}
   \includegraphics[width=1\linewidth]{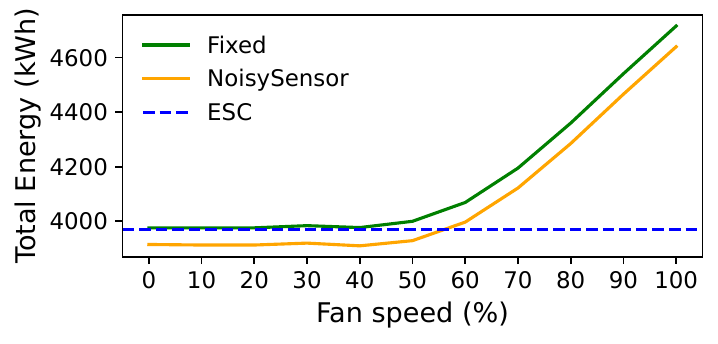}
   \caption{Pasco, WA}
   \label{fig:convex2_pasco} 
\end{subfigure}

\begin{subfigure}[b]{0.45\textwidth}
   \includegraphics[width=1\linewidth]{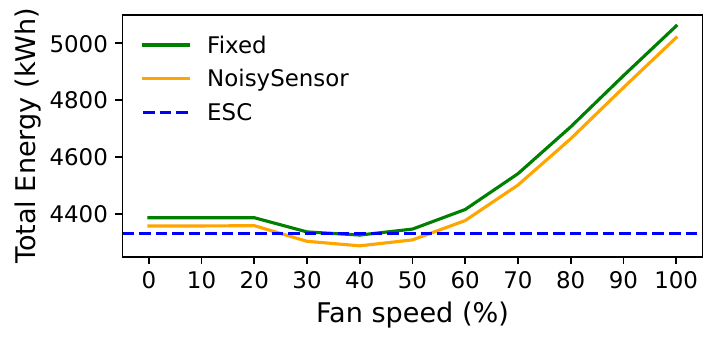}
   \caption{Houston, TX}
   \label{fig:convex2_houston}
\end{subfigure}
\caption{Convexity tests.}
\label{fig:convexity_test_noise}
\end{figure}

Figure~\ref{fig:power_stack} presents the stacked energy consumption of all components throughout the simulation testing period for both climates. Since total power consumption served as the cost function for the ESC algorithm, these plots effectively act as static maps of the cost function, indicating which fan speed minimizes total power consumption.
Figure~\ref{fig:convexity_test_noise} provides a clearer convexity test for both cases. The ``Fixed" line represents the total power consumption as fan speed varies, while the dashed line represents the total energy consumption under ESC operation. These results confirm that the ESC algorithm effectively minimized energy consumption, aligning closely with the cost function's convex minimum point.
For the Pasco Case (Figure~\ref{fig:pasco_stack}), the cumulative power was lowest at a fan speed of approximately 0–20\%. %Despite the hot climate and high cooling demand, operating the cooling tower at a low capacity suggests that it may not be energy-efficient in a dry climate like Pasco.
The lowest energy consumption with a fixed fan speed (0-20\% and 40\%) was estimated at 3975 kWh, while the energy consumption using ESC was approximately 3971 kWh, indicating better efficiency with ESC.
For Houston, the convex shape of the static map is more distinct, clearly identifying the optimal fan speed at around 40\%. This well-defined convexity suggests that the system in Houston case can operate more stably, achieving greater energy efficiency. %The analysis demonstrates that hot and humid climates, like Houston's, are particularly favorable for optimizing cooling tower fan speed, and maximizing energy savings.

% \begin{figure}[!t]
% \centering
% \begin{subfigure}[b]{0.45\textwidth}
%    \includegraphics[width=1\linewidth]{images/pasco_convex.pdf}
%    \caption{Pasco, WA}
%    \label{fig:convex2_pasco} 
% \end{subfigure}

% \begin{subfigure}[b]{0.45\textwidth}
%    \includegraphics[width=1\linewidth]{images/houston_convex.pdf}
%    \caption{Houston, TX}
%    \label{fig:convex2_houston}
% \end{subfigure}
% \caption{Convexity tests.}
% \label{fig:convexity_test}
% \end{figure}

\begin{figure}[!t]
\centering
\begin{subfigure}[b]{0.45\textwidth}
\includegraphics[width=1\linewidth]{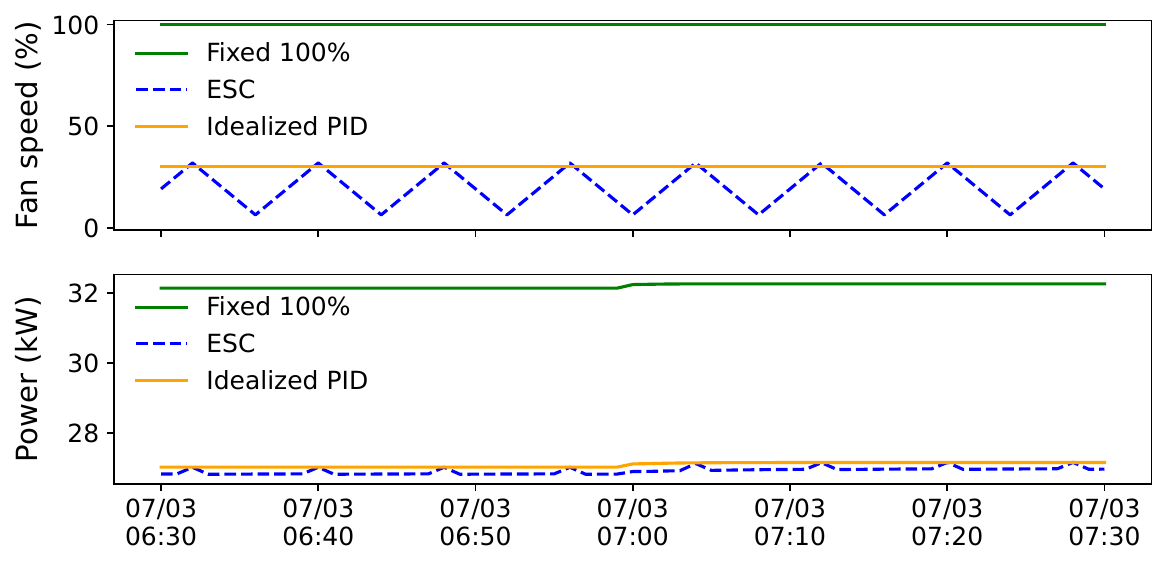}
   \caption{Pasco, WA (5B)}
   \label{fig:pasco_ESC_PID_1h} 
\end{subfigure}
\begin{subfigure}[b]{0.45\textwidth}
\includegraphics[width=1\linewidth]{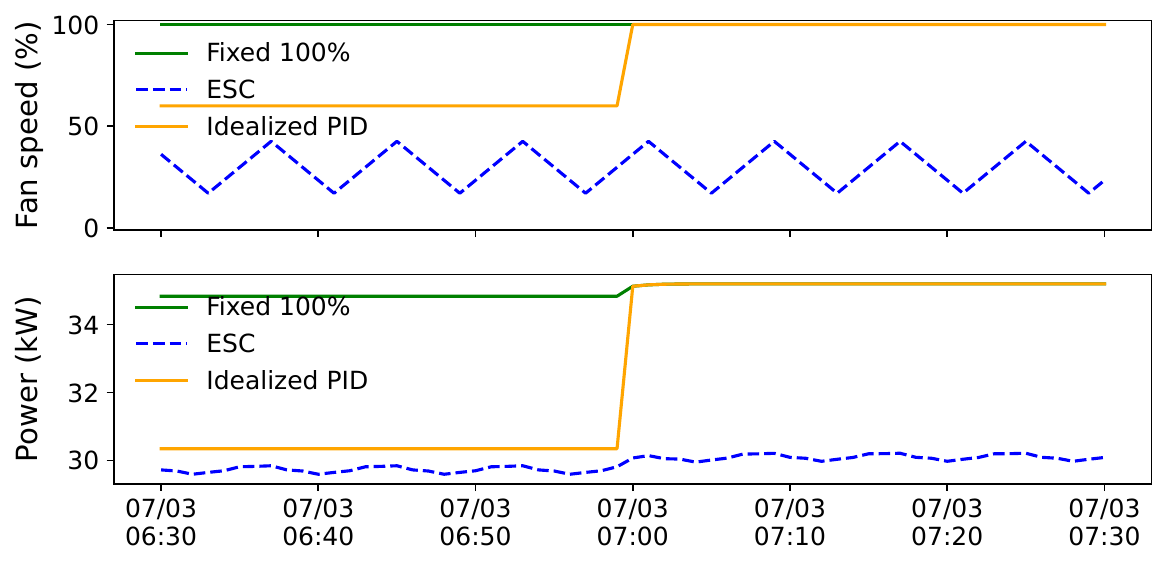}
   \caption{Houston, TX (2A)}
   \label{fig:Houston_ESC_PID_1h}
\end{subfigure}
\caption{Comparison of timeseries fanspeed and power: ESC vs Fixed 100\% vs PID}
\label{fig:fanspeed_comparison_wfixed_and_pid}
\end{figure}

\begin{figure}[!t]
\centering
\begin{subfigure}[b]{0.45\textwidth}
\includegraphics[width=1\linewidth]{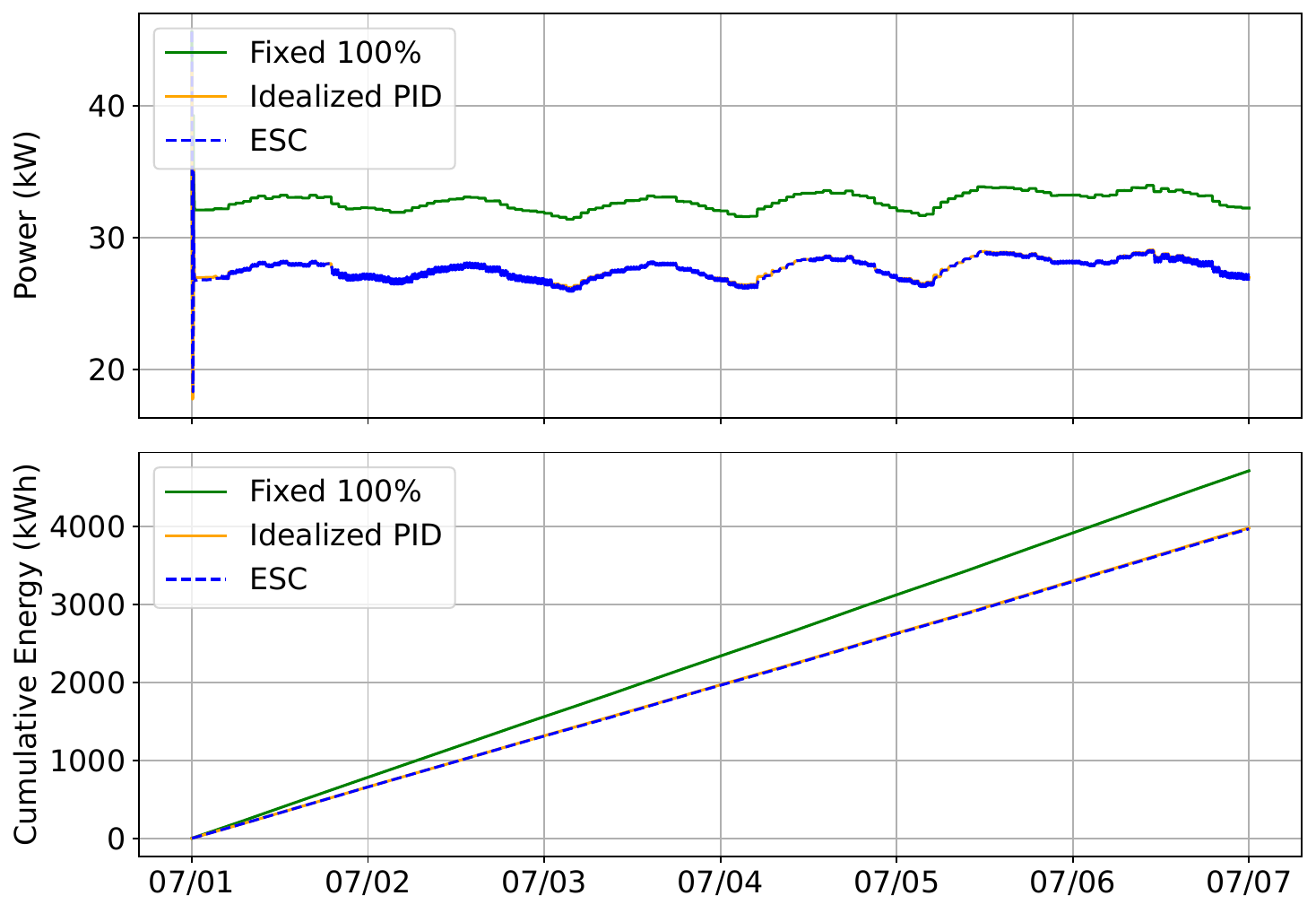}
   \caption{Pasco, WA (5B)}
   \label{fig:pasco_esc_vs_fixed100_vs_pid} 
\end{subfigure}
\begin{subfigure}[b]{0.45\textwidth}
\includegraphics[width=1\linewidth]{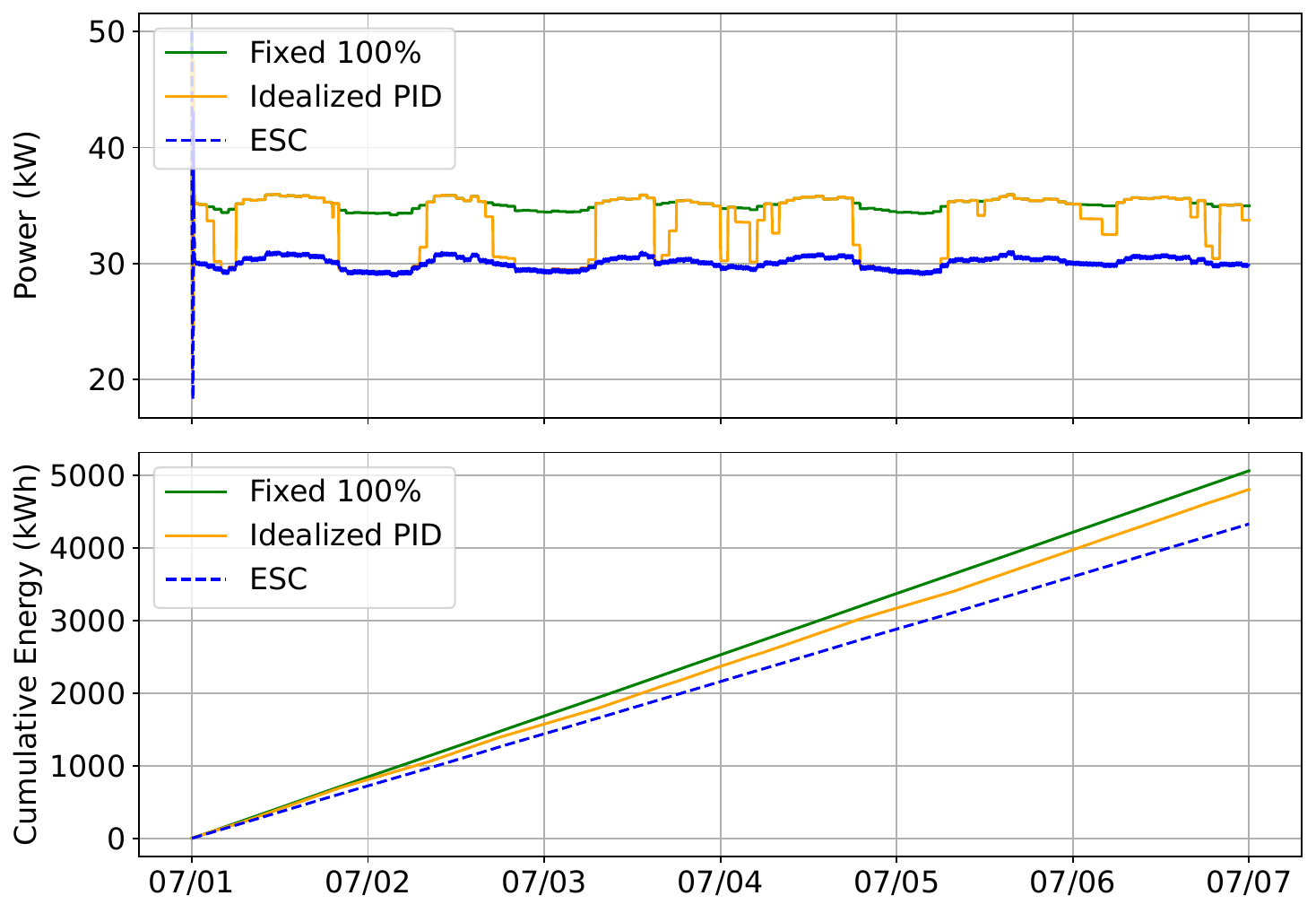}
   \caption{Houston, TX (2A)}
   \label{fig:Houston_esc_vs_fixed100_vs_pid}
\end{subfigure}
\caption{Comparison of timeseries power and cumulative energy: ESC vs Fixed 100\% vs PID}
\label{fig:power_comparison_wfixed_and_pid}
\end{figure}

\begin{figure}[!t]
\centering
\begin{subfigure}[b]{0.45\textwidth}
\includegraphics[width=1\linewidth]{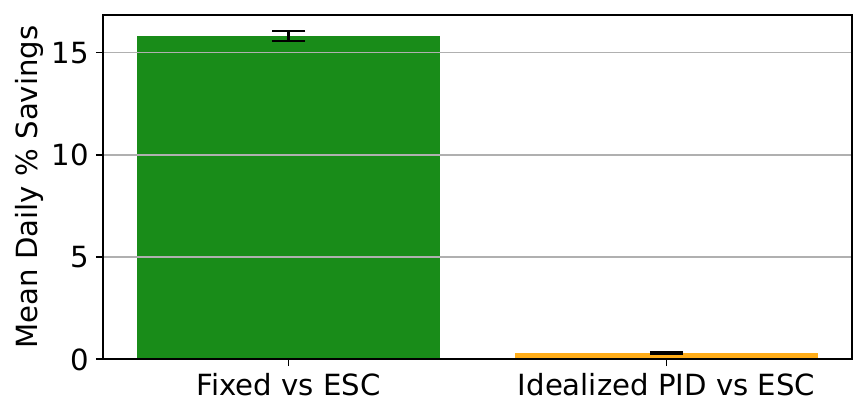}
   \caption{Pasco, WA (5B)}
   \label{fig:daily_saving_pasco} 
\end{subfigure}
\begin{subfigure}[b]{0.45\textwidth}
\includegraphics[width=1\linewidth]{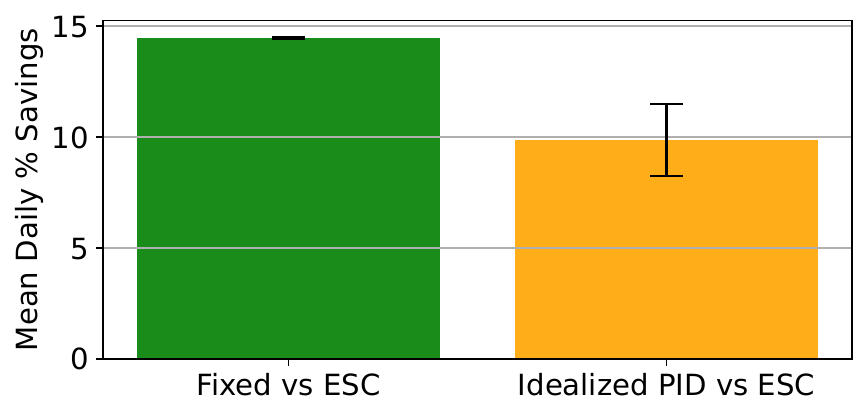}
   \caption{Houston, TX (2A)}
   \label{fig:daily_saving_Houston}
\end{subfigure}
\caption{Comparison of daily energy savings achieved by ESC relative to Fixed-speed and PID control. Bars represent the mean percentage savings, with error bars indicating 95\% confidence intervals.}
\label{fig:daily_saving}
\end{figure}

Additionally, we conducted a simple sensitivity analysis by introducing Gaussian noise (mean = 0°C, standard deviation = 5°C) to the key temperature sensors used to estimate chiller power (i.e., evaporator entering, evaporator leaving, and condenser entering temperatures). This was done to assess the impact of sensor errors on the static map. As shown by the ``NoisySensor" line in Figure~\ref{fig:convexity_test_noise}, we observed a consistent decrease in the predicted chiller power under these noisy conditions.
This effect arises from the nonlinear nature of the chiller’s power-temperature relationship. While the injected noise had zero mean, the nonlinear transformation of input temperatures can result in a biased output. For example, colder condenser entering water typically leads to reduced chiller power, while warmer chilled water return and lower supply temperatures increase the load. Adding noise distorts these relationships in unpredictable ways and it showed lower load conditions than actually exist. %In some cases, noise pushes sensor values outside the original data distribution (e.g., very low condenser temps or unusually high chilled water return), causing the chiller power to output unrealistically low power estimates. 
This behavior is consistent with Jensen’s Inequality, which states that for nonlinear functions, the expected value of the function under noisy inputs is not equal to the function evaluated at the expected inputs. Since the chiller performance curves have concave and saturating behaviors, adding noise makes the model underpredict. 
Other forms of uncertainty, such as biased measurements or missing data, could further change the original static map in complex, non-uniform ways. In general, vertical shifts in the static map (e.g., under- or over-predicted power) do not significantly impact ESC performance, as the location of the optimal operating point remains unchanged. However, horizontal shifts or deformations that affect the map's convexity can impair convergence or move the optimizer away from the true optimum.
Despite these challenges, the static map's convexity can be found even under noise, and the ESC can still converge reliably to a near-optimal operating point.

Moreover, Figure~\ref{fig:fanspeed_comparison_wfixed_and_pid} compares the time series fan speed and Figure~\ref{fig:power_comparison_wfixed_and_pid} compares time series power and cumulative energy under three control strategies: ESC, fixed 100\% control, and an idealized PID control. The PID controller in this comparison is assumed to be ideal in that the controlled variable (cooling tower leaving temperature) tracks the setpoint perfectly, which is set to a common industry standard of 25$^{\circ}$C \cite{johnson2025coolingtower}. In Pasco, which has a hot and dry climate, PID control closely mirrored ESC performance. Fan speeds remained mostly around 40\% and occasionally dropped to 0\%. These instances of 0\% fan speed do not necessarily indicate low cooling demand. They may reflect the fact that under dry ambient conditions, evaporative heat rejection in the cooling tower can be highly effective just through natural evaporation, without requiring mechanical fan operation. This behavior aligns with the ESC convexity test results, which showed minimal power between 0-40\% fan speed. In this case, PID control consumed only about 0.3\% more energy compared to ESC. In contrast, Houston's hot and humid climate posed a different aspect. To meet the same outlet temperature setpoint, the PID controller had to operate the fan at higher speeds, often near 100\%. As a result, the energy consumption with PID control was significantly higher compared to ESC, which modulated the fan speed more efficiently. In this scenario, ESC achieved approximately 9.9\% energy savings over PID. 
This comparison emphasized how ESC can maintain strong performance across diverse climate conditions by balancing energy use and cooling performance. Compared to fixed 100\% fan speed, which is a common but inefficient practice in many buildings, ESC consistently achieved 14-15\% energy savings in both locations, underscoring its practical value. Meanwhile, the PID results highlight how climate-specific conditions can affect the relative efficiency of strict setpoint-based control strategies. We also note that the performance achieved by the idealized PID control in this study would not be attainable in practice due to setpoint tracking error and the common problem of poor tuning leading to unstable behavior.
These findings align with prior research on temperature reset and dynamic fan control, reinforcing the importance of using adaptive strategies like ESC to optimize energy efficiency in cooling tower operations under varying environmental conditions.
Additionally, we estimated statistical metrics such as standard deviation, confidence intervals, and distribution characteristics as shown in Figure~\ref{fig:daily_saving}. 
In Pasco, the energy savings achieved by the ESC compared to the traditional fixed-speed control averaged 15.81\%, with a standard deviation of 0.29\%. The 95\% confidence interval (CI) ranges from 15.58\% to 16.04\%, indicating a statistically consistent improvement. This narrow CI reflects low variability and high confidence in the observed savings. When compared to the PID control strategy, the savings are much smaller, averaging just 0.30\%, with a standard deviation of 0.08\% and a 95\% CI of 0.24\% to 0.36\%. In Houston, ESC achieved an average energy savings of 14.47\% compared to fixed-speed control, with a remarkably low standard deviation of 0.08\% and a 95\% CI of 14.41\% to 14.52\%. This high degree of consistency reinforces the reliability of ESC under Houston’s climate and system conditions. Relative to PID, ESC delivered a more substantial average savings of 9.85\%, though with higher variability (standard deviation: 2.03\%, CI: 8.23\% to 11.48\%). This wider CI suggests that the benefit of ESC over PID in Houston is more sensitive to day-to-day operational or environmental factors.

\subsection{Performance of Virtual Power Meter}
% \textcolor{red}{comparison of estimated vs actual energy consumption -> virtual power meter accuracy and effectiveness in estimating energy use}

This section demonstrates the functionality of the VPM model, which is critical for implementing our ESC algorithm in scenarios where the actual building does not have a physical power meter. 
For this demonstration purpose, we focused on validating the VPM model specifically for the chiller.
We used a test building equipped with a physical power meter on its chiller system. 

\begin{table}[h!]
\centering
\begin{tabular}{|>{\raggedright\arraybackslash}m{3cm}|>{\raggedright\arraybackslash}m{3cm}|}
\hline
\textbf{Specification} & \textbf{Details} \\ \hline
Capacity & 155 tons \\ \hline
Full load efficiency & 1.2245 kW/ton \\ \hline
Part load efficiency & 0.8824 kW/ton \\ \hline
Reference COP & 3.05 \\ \hline
Compressor Type & Screw \\ \hline
Condenser Type & Air \\ \hline
Compressor Speed & Constant \\ \hline
\end{tabular}
\caption{Chiller Specifications}
\label{tab:chiller-specs}
\end{table}

\begin{figure}[!t]
\centering
\begin{subfigure}[b]{0.45\textwidth}
   \includegraphics[width=1\linewidth]{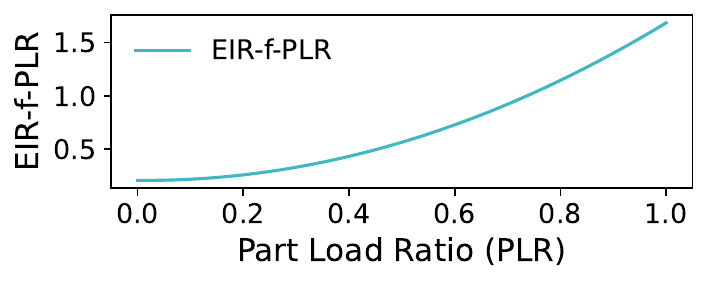}
   % \caption{eir_f_plr}
   \label{fig:convex2_pasco} 
\end{subfigure}
\begin{subfigure}[b]{0.45\textwidth}
   \includegraphics[width=1\linewidth]{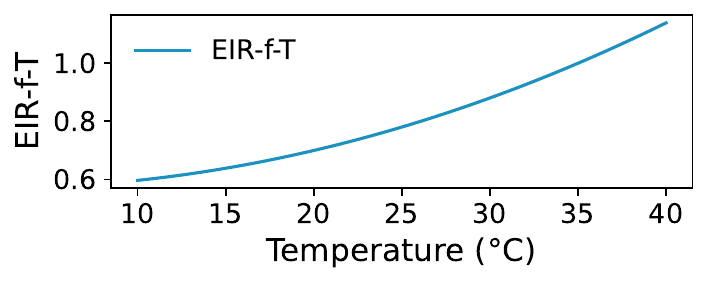}
   % \caption{eir_f_t}
   \label{fig:eir_f_t} 
\end{subfigure}
\begin{subfigure}[b]{0.45\textwidth}
   \includegraphics[width=1\linewidth]{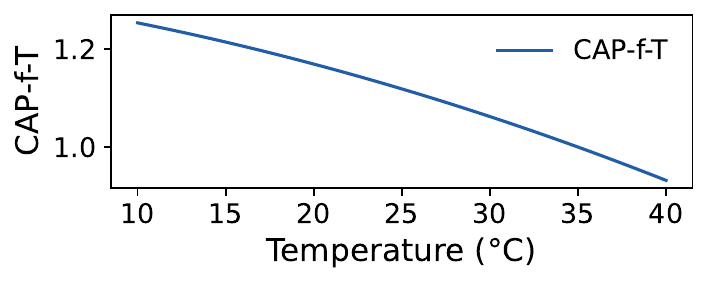}
   % \caption{cap_f_t}
   \label{fig:cap_f_t}
\end{subfigure}
\caption{Performance curves of chiller.}
\label{fig:performance_curves}
\end{figure}

To implement the VPM, we first configured its parameters. Based on the specifications of the chiller system provided in Table~\ref{tab:chiller-specs}, we generated three performance curves, as shown in Figure~\ref{fig:performance_curves}. The chiller has a capacity of 155 tons with a reference COP of 3.05. %Being air-cooled, the chiller uses outdoor air temperature for its condenser. 

For this analysis, we used historical data from May 2023 to May 2024, recorded at 15-minute intervals.
While we were able to collect most of the required measurements, chilled water flow rate data was unavailable. Instead, we assumed a constant flow rate based on the system specifications and applied a correction factor to adjust the estimated power from the VPM. Additionally, as the data was derived from 15-minute interval historical records, which may not capture smooth changes in temperature and power usage, we applied smoothing techniques to improve the analysis. % Note that discrepancies between the assumed flow and the actual flow will lead to errors in the virtual power meter estimates.

\begin{figure}[!t]
    \centering
    \includegraphics[width=0.48\textwidth]{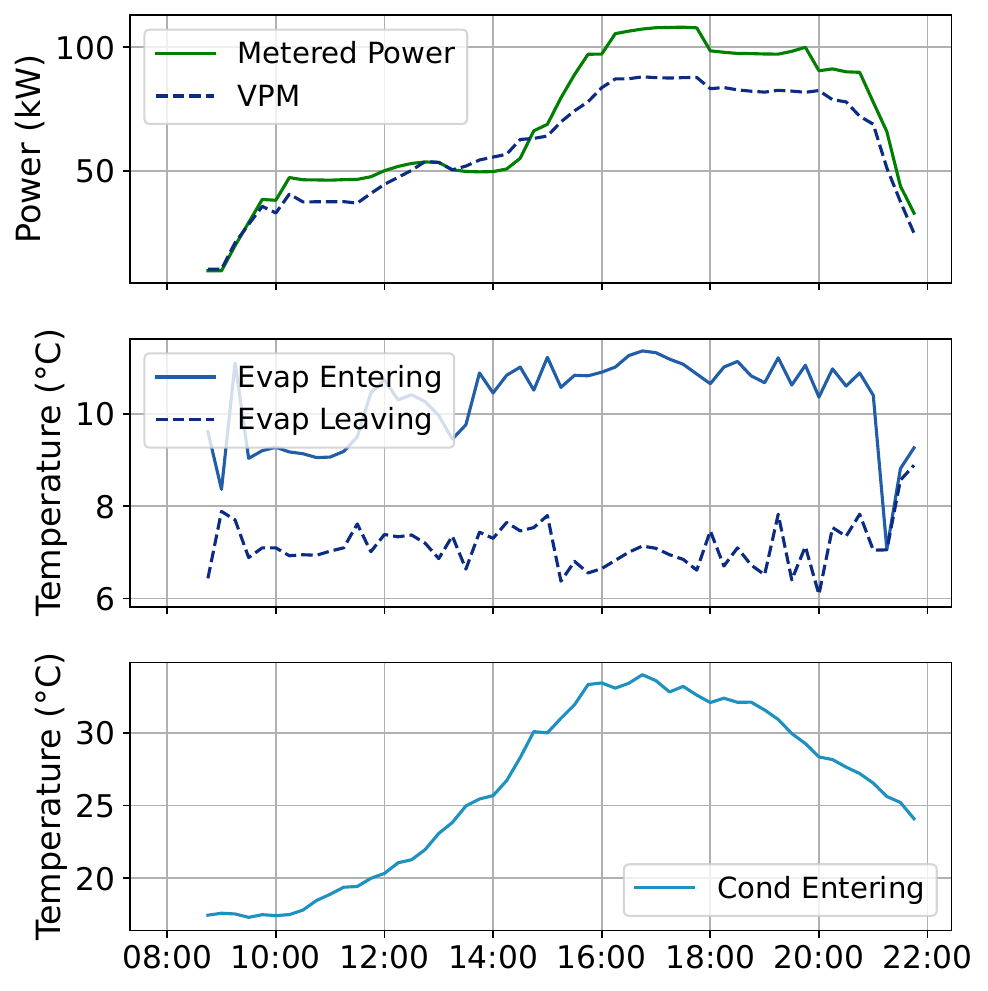}
    % \vspace{-0.15 cm}
    \caption{Virtual Power Meter performance.}
    \label{fig:vpm_performance}
    % \vspace{-0.65cm}
\end{figure}

Figure~\ref{fig:vpm_performance} presents a one-day snapshot of the chiller’s performance profile. The top subplot compares the power readings from the physical and virtual power meters. The second subplot shows the evaporator’s entering and leaving water temperatures, while the third subplot displays the condenser entering temperature, which corresponds to the outdoor air temperature, being an air-cooled chiller.% The last subplot tracks the estimated COP over time.

From the comparison of power readings, the VPM closely follows the trend observed in the physical power meter. As we assumed a constant flow rate, variations in the actual flow may have contributed to some deviations in the afternoon. Nonetheless, the VPM’s trend alignment is adequate for our purposes. 
Moreover, the ESC algorithm uses time scale separation, allowing the system to settle and overcome transient dynamics. This approach reduces the impact of short-term noise or simplifications like the fixed flow rate assumption in the VPM when data is not available. Since the ESC algorithm relies on the gradient of the cost function, capturing the correct trend direction is sufficient for effective implementation, and it supports the effective operation of ESC despite the approximation.

%COP values were higher when the condenser entering temperature was around 20°C, which aligns with the chiller’s performance curve shown in Figure~\ref{}.

\begin{figure}[!t]
    \centering
    \includegraphics[width=0.48\textwidth]{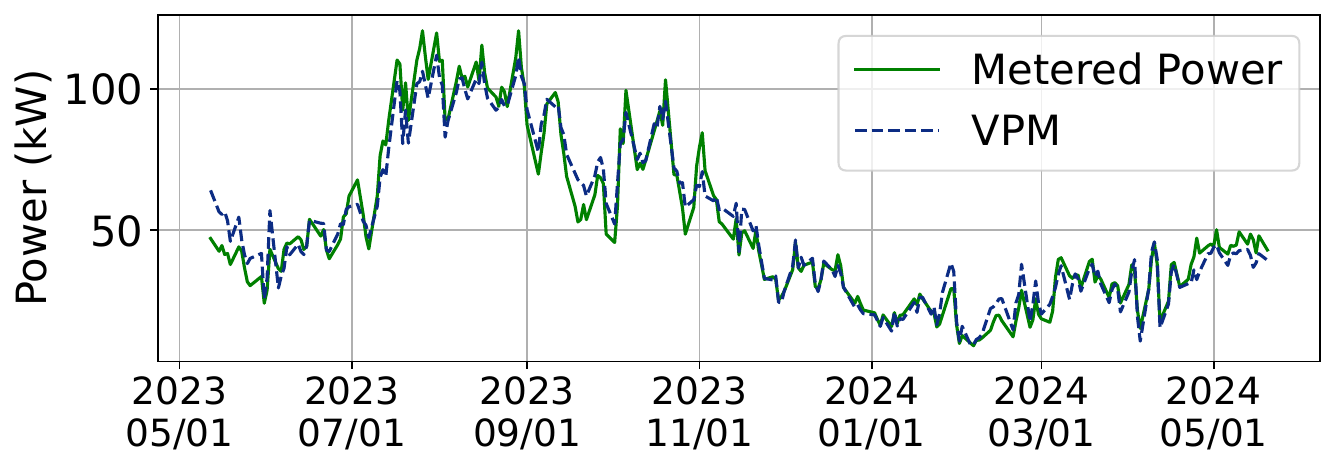}
    % \vspace{-0.15 cm}
    \caption{Comparison of physical and virtual power meter over one year.}
    \label{fig:vpm_1year}
    % \vspace{-0.65cm}
\end{figure}

\begin{figure}[!t]
    \centering
    \includegraphics[width=0.48\textwidth]{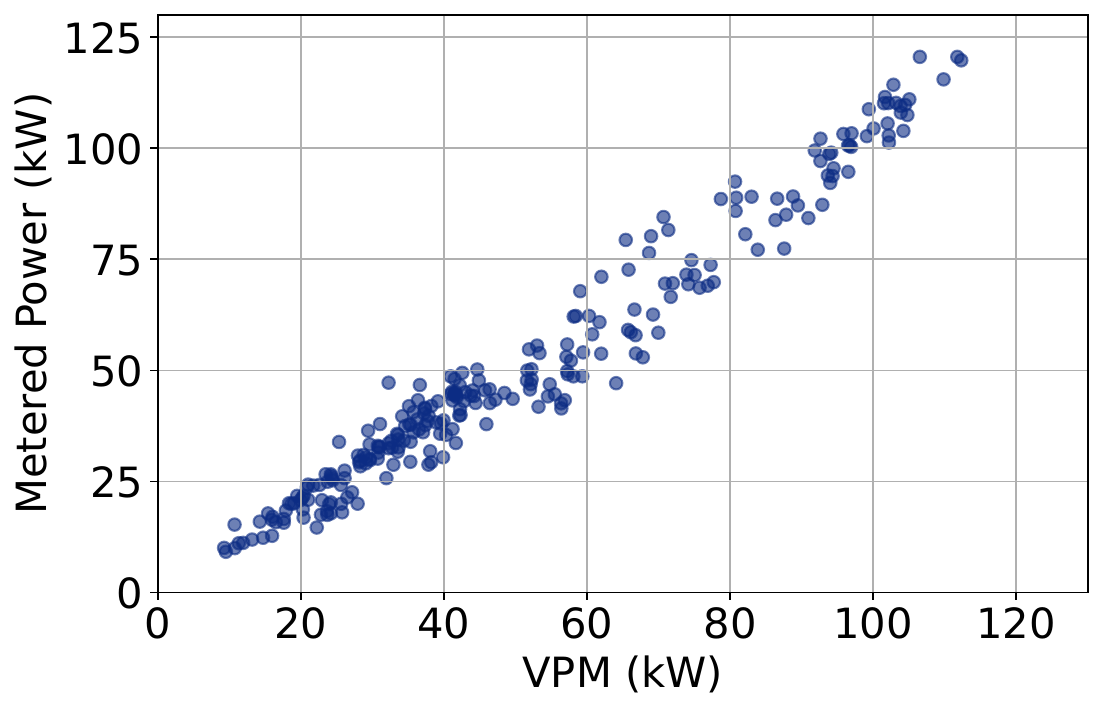}
    % \vspace{-0.15 cm}
    \caption{Scatter plot of daily average power readings (R$^2$=0.9611).}
    \label{fig:vpm_linear}
    % \vspace{-0.65cm}
\end{figure}

\begin{figure}[!t]
    \centering
    \includegraphics[width=0.48\textwidth]{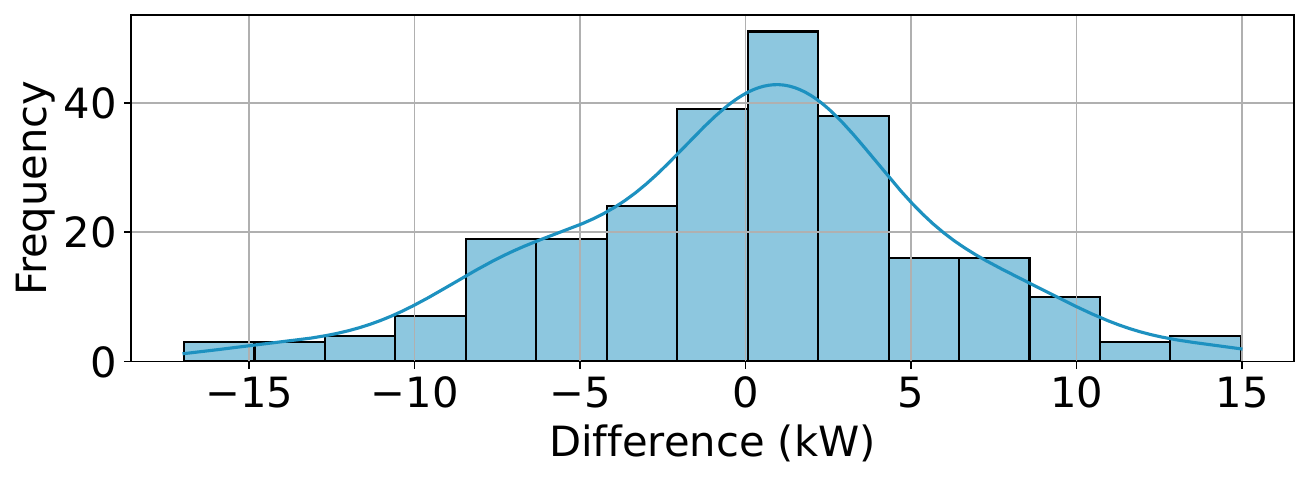}
    % \vspace{-0.15 cm}
    \caption{Histogram of differences between physical and virtual power readings.}
    \label{fig:vpm_hist}
    % \vspace{-0.65cm}
\end{figure}

Figure~\ref{fig:vpm_1year} provides a comparison of the physical and virtual power meter readings over one year. It further confirms the VPM’s ability to replicate trends, despite some minor variations due to previously discussed factors.
Figures~\ref{fig:vpm_linear} and \ref{fig:vpm_hist} provide additional validation through a scatter plot of daily average power readings and a histogram of differences between the physical and virtual power meters. The scatter plot shows that the points closely align along the diagonal, demonstrating a positive linear relationship. The R$^2$ value of 0.9611 indicates that approximately 96.11\% of the variance in metered power is explained by the VPM model, signifying a strong correlation while leaving some room for improvement in capturing variability and reducing error due to assumptions (e.g., fixed flow rates).
In addition, the RMSE and NRMSE values are 5.69 kW and 0.0511, respectively. The RMSE indicates that, on average, the virtual power meter estimation error is 5.69 kW. The NRMSE value means that the model's error is about 5.11\% relative to the range of the actual metered power data. This value is relatively low, suggesting that the VPM's accuracy is quite high.

Figure~\ref{fig:vpm_hist} highlights that the virtual power meter estimation generally aligns well with the physical measurements, as the differences are mostly small and distributed around 0. However, a deviation of up to +/- 15 kW may 
%a slight bias in the Virtual Power Meter through the difference between the physical measurement and the VPM values. The mean difference is approximately -10 kW, suggesting a tendency for the VPM to slightly overestimate power. The spread in the histogram 
reflect factors such as missing data (e.g., flow rate), so that the use of a fixed constant flow rate with simplifying assumptions in the VPM setup impacts the accuracy.

Overall, the results demonstrate that the VPM is functional and reliable enough for use in buildings and systems lacking physical power meters. The VPM effectively captures general power trends, making it a suitable tool for approximating metered power. While the high R$^2$ value indicates a reliable model, %points to a reliable model, 
further calibration or incorporation of additional variables could enhance accuracy. Importantly, the VPM's performance is sufficient to enable the application of the ESC algorithm on chilled water plant systems effectively.

\section{Conclusion} \label{sec:conc}
This paper introduced an innovative control method for cooling tower operation designed to minimize the overall power consumption of a chilled water plant system. The proposed approach utilizes a relay-based ESC algorithm for real-time optimization. This paper demonstrated its effectiveness in different climate locations through a simulation-based study, where a detailed simulation model of a chilled water plant served as the testbed. The control system was developed and implemented in Python, with a modular co-simulation framework to integrate the controller and system model seamlessly.

The ESC-based real-time optimizer dynamically adjusts the condenser water loop to track the optimal trade-off between chiller and cooling tower power in real-time. Simulation results highlighted potential energy savings of approximately 15\% by optimizing cooling tower fan speed compared to operating fans at fixed speed (100\%) during summer conditions.

Building on the simulation results, we developed a VPM to estimate power consumption in real-world systems using sensor data. This approach addresses the challenge of implementing ESC in real-world systems that lack physical power meters. 
Among our test buildings, only one had a physical power meter installed on the chiller system. 
This not only allowed us to validate the VPM's performance against physical meter measurements but also highlighted the importance of the development of the VPM. Our performance comparison study demonstrated a strong correlation of 96.11\% even with some measurements missing (flow rate). Additionally, a 5.11\% normalized error suggests that, for the real-world applications, the virtual power meter estimations can be considered highly reliable. As our ESC approach utilizes the gradient of the cost function, its reliable trend for the power consumption of the VPM supports the ESC's functionality in energy optimization, without physical power meters. 

While the ultimate goal is to implement this control system in real-world applications, this paper focused on simulation-based validation of the ESC-based real-time optimizer and the demonstration of the VPM's capabilities. Our future work will be implementing this real-time optimizer in the real-world chilled water plant system to validate the system's effectiveness in practical scenarios. We also plan to enhance the robustness of our VPM to account for scenarios with missing data and to further explore uncertainty analysis associated with the ESC performance. % and when sensor measurements are available at longer intervals (e.g., 10-15 minutes) instead of 1-minute intervals. This will involve capturing and addressing the regression effects caused by the extended time intervals in sensor measurements.
 As the proposed approach is entirely software-based, it offers a cost-effective and scalable solution for broader adoption across a diverse range of chilled water system types.

\section{Acknowledgment}
This work is supported by the U.S. Army Reserve Installation Management Directorate (ARIMD) Enterprise Building Control System (EBCS) and the Environmental Security Technology Certification Program (ESTCP).

\section{Data Availability Statement}
The data supporting the findings of this study are available from the corresponding author upon reasonable request.

\section{Disclosure statement}
No potential conflict of interest was reported by the author(s).

% % --------------------------------------------------------
% \section{Nomenclature}

% \begin{tabular}{p{12mm}p{55mm}}
%   $e$        & error \\
%   $E$        & energy \\
%   $T$        & temperature \\
%   $\epsilon$ & solver precision parameter \\
%   $\epsilon^*$ & highest setting for solver precision parameter \\
%   $a \in A$      & $a$ is an element of $A$\\
%   $\Re$      & set of real numbers \\
%   $\triangleq$ & equal by definition \\
% \end{tabular}

% --------------------------------------------------------

\bibliographystyle{achicago}
\bibliography{xbib}

\end{document}